\documentclass[aps,prb,twocolumn,superscriptaddress]{revtex4-2}
\usepackage{graphics,graphicx, array, afterpage}
\usepackage{amsmath}
\usepackage{grffile}
\usepackage[export]{adjustbox}
\usepackage{lineno}
\usepackage{xcolor}
\usepackage{longtable}
\usepackage{braket}
\usepackage{array}
\usepackage{multirow}
\usepackage{enumerate}
\usepackage{amsmath}
\usepackage{amssymb}
\usepackage{amsthm}
\usepackage{times}
\usepackage{multirow}
\usepackage[utf8]{inputenc}
\usepackage[english]{babel}
\usepackage[T1]{fontenc}
\usepackage{amsmath,amssymb,dsfont}
\usepackage[colorlinks,
citecolor=blue,
linkcolor=blue,
urlcolor=blue]{hyperref}
\usepackage{tikz}
\usepackage{lipsum}
\usepackage{dsfont}
\usepackage{physics}
\usepackage{float}
\usepackage{braket}
\usepackage{orcidlink}
\usepackage{multirow}
\usepackage{algorithm}
\usepackage{algpseudocode}
\usepackage{graphbox}
\usepackage{amsfonts}
\usepackage{dcolumn}
\usepackage{epstopdf}
\usepackage{lipsum}
\usepackage{amsmath,epsfig,amssymb,subfigure,bm,dsfont}
\begin{document}

\title{Slow growth of quantum magic in disorder-free Stark many-body localization}

\author{Han-Ze Li}
\email{hanzeli@u.nus.edu}
\affiliation{Institute for Quantum Science and Technology, Shanghai University, Shanghai 200444, China}
\affiliation{Department of Physics, National University of Singapore, Singapore 117542, Singapore}

\author{Yi-Rui Zhang}
\affiliation{Institute for Quantum Science and Technology, Shanghai University, Shanghai 200444, China}

\author{Yu-Jun Zhao}
\affiliation{Institute for Quantum Science and Technology, Shanghai University, Shanghai 200444, China}
\affiliation{School of Physics and Optoelectronics, Xiangtan University, Xiangtan 411105, China}

\author{Xuyang Huang}
\email{hxy_@shu.edu.cn}
\affiliation{Institute for Quantum Science and Technology, Shanghai University, Shanghai 200444, China}

\author{Jian-Xin Zhong}
\email{jxzhong@shu.edu.cn}
\affiliation{Institute for Quantum Science and Technology, Shanghai University, Shanghai 200444, China}
\affiliation{School of Physics and Optoelectronics, Xiangtan University, Xiangtan 411105, China}

\begin{abstract}
Disorder-free quantum many-body localization can strongly suppress transport while still enabling the dynamical buildup of computationally costly non-Clifford resources. In a tilted transverse-field Ising chain realizing disorder-free Stark many-body localization, we use the stabilizer R\'enyi entropy to quantify quantum magic (nonstabilizerness) and find that it remains finite and grows anomalously slowly over extended time windows before saturating to a size-dependent plateau deep in the strong-tilt regime, with pronounced initial-state selectivity. Upon increasing the Stark gradient, the long-time magic and half-chain entanglement exhibit consistent finite-size crossing behavior, indicating a crossover from ergodic dynamics to constrained localization. These results establish stabilizer-based magic as a practical complexity diagnostic of disorder-free ergodicity breaking and constrained dynamics, and provide an experimentally accessible route to benchmarking and designing near-term quantum simulators.
\end{abstract}

\maketitle
\section{Introduction}
The quest to identify the distinct quantum resources that enable computational advantage over classical devices lies at the heart of modern quantum information science~\cite{Feynman1982Simulating,Lloyd1996UniversalSimulators,Shor1997Algorithms,Grover1997Search,BernsteinVazirani1997QCT,AaronsonArkhipov2011LinearOptics,Zhong2020PhotonicAdvantage,Kretschmer2025}. While multipartite entanglement is a necessary condition for quantum speedup, it is not sufficient: according to the Gottesman--Knill theorem~\cite{gottesman1998heisenbergrepresentationquantumcomputers,PhysRevA.70.052328}, states generated by Clifford circuits can be highly entangled yet efficiently simulated on classical computers in polynomial time. To capture the true cost of quantum simulation and the potential for universality, one must quantify \emph{quantum magic} (nonstabilizerness)~\cite{chitambar2019quantum,liu2022manybody,Leone_2022,Leone_2024,tirrito2024quantifying,Turkeshi_2023,turkeshi2025pauli,Turkeshi_2024_2,jasser2025stabilizerentropyentanglementcomplexity,viscardi2025interplayentanglementstructuresstabilizer,Iannotti2025entanglement,cusumano2025nonstabilizernessviolationschshinequalities,bittel2025operationalinterpretationstabilizerentropy,varikuti2025impactcliffordoperationsnonstabilizing,tirrito2025anticoncentrationnonstabilizernessspreadingergodic,zhang2025stabilizerrenyientropytransition,qian2025quantum,moca2025nonstabilizernessgenerationmultiparticlequantum,dowling2025magic,bera2025nonstabilizernesssachdevyekitaevmodel,masotllima2024stabilizer,aditya2025mpembaeffectsquantumcomplexity,hernándezyanes2025nonstabilizernessquantumenhancedmetrologicalprotocols,falcão2025magicdynamicsmanybodylocalized,sticlet2025nonstabilizernessopenxxzspin,tirrito2025universalspreadingnonstabilizernessquantum,zhang2024quantummagicdynamicsrandom,cao2025gravitationalbackreactionmagical,Tarabunga_2024,qiant25,qian2025,hang2025,qiant2024a,frau2025,fan2025disentangling,huang2024cliffordcircuitsaugmentedmatrix,ding2025evaluating,korbany2025longrangenonstabilizernessphasesmatter,tarabunga2025efficientmutualmagicmagic,szombathy2025independentstabilizerrenyientropy,hou2025stabilizerentanglementenhancesmagic,hoshino2025stabilizerrenyientropyconformal,tarabunga2024magictransitionmeasurementonlycircuits,tirrito2025magicphasetransitionsmonitored,wang2025magictransitionmonitoredfree,santra2025complexitytransitionschaoticquantum,Haug2025probingquantum,Haug2023stabilizerentropies,Haug_2023_1,Lami_2023_2,Lami_2024,tarabunga2023manybody,tarabunga2024nonstabilizerness,aditya2025growthspreadingquantumresources}---the resource required to implement non-Clifford gates or to prepare states outside the stabilizer manifold. Recent advances have established magic not merely as a static resource for fault-tolerant computing~\cite{Preskill2018NISQ,Quantinuum2025,Aasen2025,Peham2025}, but also as a dynamical probe for many-body physics, intimately linked to quantum chaos~\cite{LarkinOvchinnikov1969JETP,ShenkerStanford2014JHEP,MaldacenaShenkerStanford2016JHEP,Swingle2018NatPhys,HashimotoMurataYoshii2017JHEP,RobertsYoshida2017JHEP,CotlerHunterJonesLiuYoshida2017JHEP,LeoneOlivieroZhouHamma2021Quantum,GotoNosakaNozaki2022PRD,PaviglianitiLamiColluraSilva2025PRXQ,Prosen2021Chaos,HuangLiHuseChan2023Scholarpedia,li2025quantummpembaeffectlongranged}, information scrambling~\cite{HaydenPreskill2007JHEP,SekinoSusskind2008JHEP,HosurQiRobertsYoshida2016JHEP,vonKeyserlingkRakovszkyPollmannSondhi2018PRX,KhemaniVishwanathHuse2018PRX,Swingle2018NatPhys,LiuZhangYinZhang2024PRL,LiuZhangYinZhangYao2024arXiv,YuLiZhang2025CPL,HuangLiHuseChan2023Scholarpedia,PhysRevResearch.7.L012011}, and the emergence of unitary $k$-designs~\cite{DankertCleveEmersonLivine2006arXiv,DankertCleveEmersonLivine2009PRA,HarrowLow2009CMP,BrandaoHarrowHorodecki2016CMP,HoChoi2022PRL,IppolitiHo2023PRXQ,ClaeysLamacraft2022Quantum,ClaeysLamacraftVicary2024JPhysA,bnld-2chd}. Understanding how magic evolves under Hamiltonian dynamics is therefore crucial for characterizing the inherent computational complexity of quantum matter.

In parallel, the study of ergodicity breaking and the breakdown of thermalization in isolated quantum systems has identified many-body localization (MBL) as an exceptionally robust mechanism~\cite{Abanin2019RMP,NandkishoreHuse2015,AletLaflorencie2018,sxz1,sxz2,sxz3,sxz4,lhz1,lhz2}. While early work focused on disorder-driven systems~\cite{Abanin2019RMP}, it is now clear that disorder is not a prerequisite for localization: robust ergodicity breaking can also emerge in disorder-free systems governed by strong kinetic constraints~\cite{Sala2020PRX,Khemani2020PRB,Scherg2021NatCommun,Brenes2018PRL,Karpov2021PRL}. A prime example is \emph{Stark many-body localization} (SMBL), which arises in interacting lattice models subject to a strong linear potential gradient~\cite{Schulz2019PRL}. In this regime, energy conservation severely constrains transport: hopping processes that change the center of mass are resonantly suppressed, leading to a mechanism rooted in the Wannier--Stark ladder~\cite{Gluck2002PhysRep,Scherg2021NatCommun,Nandy2024PRB,lhz3,lhz4,lhz5,lhz6}. In the strong-tilt limit, these constraints can induce Hilbert-space fragmentation~\cite{Moudgalya_2022}, where the accessible Hilbert space shatters into exponentially many disconnected dynamical sectors, effectively arresting transport~\cite{Sala2020PRX,Khemani2020PRB,Scherg2021NatCommun,Adler2024Nature}. This places SMBL at the intersection of localization, constrained dynamics, and non-thermal phases in clean quantum matter~\cite{Schulz2019PRL,Sala2020PRX,Khemani2020PRB}.

The behavior of magic in such nonergodic regimes presents a fundamental puzzle. In ergodic systems, magic typically grows rapidly as information scrambles into complex Pauli strings~\cite{Nahum2018,Leone_2022}, and can approach values characteristic of Haar-random states~\cite{liu2022manybody,Odavic2025PRBStabilizerEntropy}. Conversely, in MBL phases, the dynamics are restricted by emergent quasi-local integrals of motion~\cite{Abanin2019RMP,Serbyn2013LIOM}. Recent studies have begun to chart this ``complexity frontier'' in \emph{disordered} MBL systems~\cite{Falcao2025MBLMagic,Cao2025QuantumLowWeightSRE}, suggesting that magic can provide a sensitive diagnostic across the ergodic--MBL boundary. However, disordered systems are intrinsically heterogeneous: rare Griffiths effects and thermal inclusions can dominate long-time behavior~\cite{Agarwal2017RareRegions,DeRoeckHuveneers2017Stability,Thiery2018Avalanche}, obscuring the intrinsic properties of the localized phase. More fundamentally, Stark localization is driven not by randomness but by kinetic constraints and fragmentation~\cite{Scherg2021NatCommun,Schulz2019PRL,Sala2020PRXFragmentation}, leading to a distinct dynamical structure. This raises a sharp question: does fragmentation in SMBL effectively \emph{Cliffordize} the dynamics and bound magic, or can higher-order resonant processes~\cite{Scherg2021NatCommun,vanNieuwenburg2019PNAS} still generate non-Clifford resources, albeit slowly?

In this work, we study the dynamical generation of magic in a clean constrained system exhibiting SMBL. Focusing on a tilted transverse-field Ising chain, we quantify complexity using the second stabilizer R\'enyi entropy (2-SRE) and compare it with the growth of half-chain entanglement across several global quenches from representative product states. We find that deep in the strong-tilt regime, magic continues to increase over extended time windows before saturating to a finite-size plateau. Still, the growth is strongly slowed and becomes highly initial-state-dependent. As the Stark gradient is increased, long-time magic and entanglement exhibit consistent finite-size crossing behavior, indicating a crossover from ergodic dynamics to constrained localization. In the strong-tilt limit, we further derive an effective diagonal description via a Schrieffer--Wolff (SW) expansion, where emergent diagonal couplings are factorially suppressed with distance along the Wannier--Stark ladder~\cite{PhysRev.117.432}, providing a natural explanation for the slow magic growth. Finally, we outline a trapped-ion digital-simulation protocol in which both $M_2(t)$ and second R\'enyi entanglement can be extracted from the same local randomized-measurement data, enabling an experimentally accessible characterization of slow magic dynamics.

The remainder of this paper is organized as follows. Section~\ref{sec2} introduces the model and diagnostics. Section~\ref{sec3} presents the strong-tilt analytical theory. Section~\ref{sec4} reports numerical results for the dynamics and scaling of $M_2(t)$ and its relation to entanglement, including the ergodic--SMBL analysis. Section~\ref{sec5} discusses an experimental implementation, and Sec.~\ref{sec6} concludes.

\section{Model and diagnostics}\label{sec2}

\subsection{Model}
We consider a one-dimensional chain of $L$ qubits with open boundary conditions (OBC). Sites are indexed by $i=0,1,\dots,L-1$, with Pauli operators $X_i$, $Y_i$, $Z_i$ and identity $I_i$ acting on site $i$. A tilted transverse-field Ising Hamiltonian governs the dynamics,
\begin{align}
H
&=
J \sum_{i=0}^{L-2} Z_i Z_{i+1}
+
h \sum_{i=0}^{L-1} X_i
+
F \sum_{i=0}^{L-1} i\, Z_i .
\label{eq:H_stark_tfim}
\end{align}
Here, $J$ is the nearest-neighbor Ising coupling, $h$ is the transverse-field strength, and $F$ is the Stark-gradient parameter that induces a site-dependent longitudinal field increasing linearly along the chain. We adopt OBC throughout, since the linear tilt explicitly breaks translation invariance.

\subsection{Diagnostics}
We characterize the dynamics using two complementary diagnostics. Bipartite entanglement quantifies quantum correlations across a spatial cut. At the same time, the stabilizer R\'enyi entropy (SRE) directly probes quantum magic (nonstabilizerness), i.e., how far the evolving state lies outside the stabilizer manifold and therefore beyond efficient Clifford simulation.

For a pure state $|\Psi\rangle$ of the full chain and a subsystem $R$ with complement $R^c$, the reduced density matrix is $\rho_R := \Tr_{R^c}\!\left(|\Psi\rangle\langle\Psi|\right)$. The R\'enyi entanglement entropy~\cite{Entanglement1,Entanglement2} of order $\alpha$ is
$S_\alpha(\rho_R) := \frac{1}{1-\alpha}\log_2 \Tr\!\left(\rho_R^\alpha\right)$
for $\alpha\in(0,1)\cup(1,\infty)$, and probes the full eigenvalue spectrum of $\rho_R$, with larger $\alpha$ emphasizing the largest eigenvalues more strongly. The von Neumann entropy is obtained in the limit $\alpha\to 1$,
\begin{align}
S_1(\rho_R) := -\Tr\!\left(\rho_R\log_2\rho_R\right).
\label{eq:vonNeumann_def}
\end{align}
In the main text, we focus on the half-chain bipartition $|R|=L/2$ and report the half-chain entanglement entropy $S_{L/2}$ as a standard indicator distinguishing ergodic from constrained dynamics.

Entanglement alone does not fully capture computational complexity, since stabilizer states can be highly entangled yet remain classically efficiently simulable. To directly diagnose magic, we employ the stabilizer R\'enyi entropy~\cite{Leone_2022,Leone_2024}. Let $\mathcal{P}_L \!:=\! \{I,X,Y,Z\}^{\otimes L}$ denote the set of $L$-qubit Pauli strings up to overall phases, and let $D:=2^L$ be the Hilbert-space dimension. For a pure state $|\Psi\rangle$ we define the stabilizer purity
\begin{align}
P_\alpha(\Psi)
:=
\frac{1}{D}\sum_{P\in\mathcal{P}_L}\big|\langle\Psi|P|\Psi\rangle\big|^{2\alpha},
\end{align}
and the SRE
\begin{align}
M_\alpha(\Psi)
:=
\frac{1}{1-\alpha}\log_2 P_\alpha(\Psi),
\quad
\alpha>0,\ \alpha\neq 1,
\label{eq:SRE_def}
\end{align}
with $M_1$ defined by the limit $\alpha\to 1$. Throughout we fix $\alpha=2$ and denote $M_2\equiv M_{\alpha=2}$ (the 2-SRE). With this convention, $M_2\ge 0$, and $M_2=0$ holds if and only if $|\Psi\rangle$ is a pure stabilizer state. Moreover, $M_2$ is invariant under Clifford unitaries, since Cliffords permute Pauli strings under conjugation and preserve the multiset of Pauli moments, and it is additive under tensor products:
$M_2(\Psi\otimes\Phi)=M_2(\Psi)+M_2(\Phi)$.
Consequently, any increase of $M_2$ under unitary time evolution witnesses the generation of non-Clifford resources. Importantly, SRE is experimentally accessible via randomized measurement protocols~\cite{Oliviero_2022,PRXQuantum.4.010301,Niroula2024}.

As a reference for highly complex states, we compare $M_2$ to its Haar-random value~\cite{Leone_2022,turkeshi2025pauli,Turkeshi_2024_2,Odavic2025PRBStabilizerEntropy}. The expected 2-SRE of Haar-random pure states is
\begin{align}
M_2^{\mathrm{Haar}}
=
\log_2(2^L+3)-2,
\label{eq:M2_haar}
\end{align}
which approaches $L-2$ at large $L$. Reaching this benchmark is a dynamic question whose answer depends on the structure of the unitary evolution. In random unitary circuits~\cite{turkeshi2025pauli,Turkeshi_2024_2,tirrito2025anticoncentrationnonstabilizernessspreadingergodic,tirrito2025universalspreadingnonstabilizernessquantum,Turkeshi_2023,Turkeshi_2024_2}, local randomness is injected layer by layer and $M_2$ typically approaches $M_2^{\mathrm{Haar}}$ at depths that grow only slowly with $L$~\cite{Odavic2025PRBStabilizerEntropy}. Generic Floquet dynamics without conservation laws can display similarly fast, circuit-like scrambling~\cite{DankertCleveEmersonLivine2009PRA,HarrowLow2009CMP,BrandaoHarrowHorodecki2016CMP}. In contrast, for local Hamiltonian evolution the approach to the Haar benchmark is constrained by locality: operator support and information must propagate across the chain, implying an approach time that grows at least linearly with $L$, while accessible-time saturation may remain below $M_2^{\mathrm{Haar}}$ due to residual structure from locality, symmetries, and conserved quantities. These considerations are particularly relevant here, where the linear potential further restricts resonant processes and can strongly suppress the generation of magic. For convenience, we also use the deviation from the Haar benchmark~\cite{falcão2025magicdynamicsmanybodylocalized},
\begin{align}
\Delta M_2 := M_2^{\mathrm{Haar}}-M_2,\label{eq:deltaM2}
\end{align}
so that ergodic dynamics correspond to small $\Delta M_2$, while constrained or nonergodic regimes exhibit a parametrically larger deviation.

We investigate global quenches $|\Psi(t)\rangle\!=\!e^{-iHt}|\Psi_0\rangle$ using four families of initial product states:
(i) a specific computational-basis state $|z_\star\rangle$ chosen to minimize the diagonal energy $|\langle z|H_{\mathrm{diag}}|z\rangle|$;
(ii) the fully $X$-polarized state $|\Psi_X\rangle\!=\!|+\rangle^{\otimes L}$ with $|+\rangle=\frac{1}{\sqrt2}(|0\rangle+|1\rangle)$;
(iii) the fully $Y$-polarized state $|\Psi_Y\rangle\!=\!|+_y\rangle^{\otimes L}$ with $|+_y\rangle\!=\!\frac{1}{\sqrt2}(|0\rangle+i|1\rangle)$; and
(iv) an ensemble of Haar-random product states, over which we average the observables.

\section{Analytical results}\label{sec3}
To capture the slow magic dynamics in the strong-tilt regime, we derive an effective (predominantly) diagonal description by perturbatively integrating out off-resonant transverse-field processes. In the limit $F\!\gg\! J,h$, we split Eq.~\eqref{eq:H_stark_tfim} as $H\!=\!H_0+V$, where (consistent with the site indexing $j\!=\!0,1,\dots,L-1$ and OBC used throughout)
\begin{align}
H_0 &= J\sum_{j=0}^{L-2} Z_j Z_{j+1} + F\sum_{j=0}^{L-1} j\,Z_j,
\quad
V = h\sum_{j=0}^{L-1} X_j .
\end{align}
We perform a Schrieffer--Wolff (SW) transformation $H_{\rm eff}\!=\!e^{S}He^{-S}$ with an anti-Hermitian generator $S^\dagger\!=\!-S$, choosing $S^{(1)}=O(h)$ such that $[H_0,S^{(1)}]\!=\!V$, which removes the leading off-diagonal processes in the eigenbasis of $H_0$. Truncating consistently yields $H_{\rm eff}\!=\!H_0+\tfrac12[S^{(1)},V]+O(h^3)$, while higher SW orders generate diagonal multi-spin $Z$-string couplings. Since long-time dynamics in the strong-tilt regime is dominated by dephasing generated by such diagonal terms, we retain only the diagonal sector $H_{\rm eff}^{(d)}$ and quantify the strength of diagonal two-body couplings at separation $r$ by a typical scale $J_{\rm eff}(r)$ (see Appendix~\ref{sec:Analytical-details} for explicit definitions and derivations).

In the strong-tilt regime, intermediate-state energy denominators accumulate along the Stark ladder, leading to a factorial suppression with distance,
\begin{align}
J_{\rm eff}(r) \sim J_0\,\frac{(h/F)^{r-1}}{(r-1)!},
\label{eq:Jeff_factorial}
\end{align}
where $J_0\sim h$ sets the microscopic scale. In the dephasing-dominated regime $tJ_0\gtrsim 1$, we define a dephasing front $r(t)$ by the condition $tJ_{\rm eff}(r(t))\sim 1$, which gives the closed-form estimate
\begin{align}
r(t) \simeq 1 + \frac{\ln(tJ_0)}{W_0\!\left(\dfrac{\ln(tJ_0)}{e\,h/F}\right)},
\label{eq:r_of_t}
\end{align}
where $W_0$ is the principal branch of the Lambert $W$ function defined by $W_0(x)e^{W_0(x)}\!=\!x$. Motivated by an extensive growth controlled by the active length scale $r(t)$ and eventual finite-size saturation, we use the following compact saturating closure for the 2-SRE,
\begin{align}
M_2(t) \simeq M_2^{\rm sat}\,\tanh\!\big(\gamma\, r(t)\big),
\label{eq:log_growth}
\end{align}
where $M_2^{\rm sat}$ is the finite-size plateau and $\gamma$ is a (generally $F$- and initial-state-dependent) growth coefficient. Full derivations are provided in Appendix~\ref{sec:Analytical-details}.

\section{Numerical results}\label{sec4}

\begin{figure}[bt]
\hspace*{-0.47\textwidth}
\centering
\includegraphics[width=0.47\textwidth]{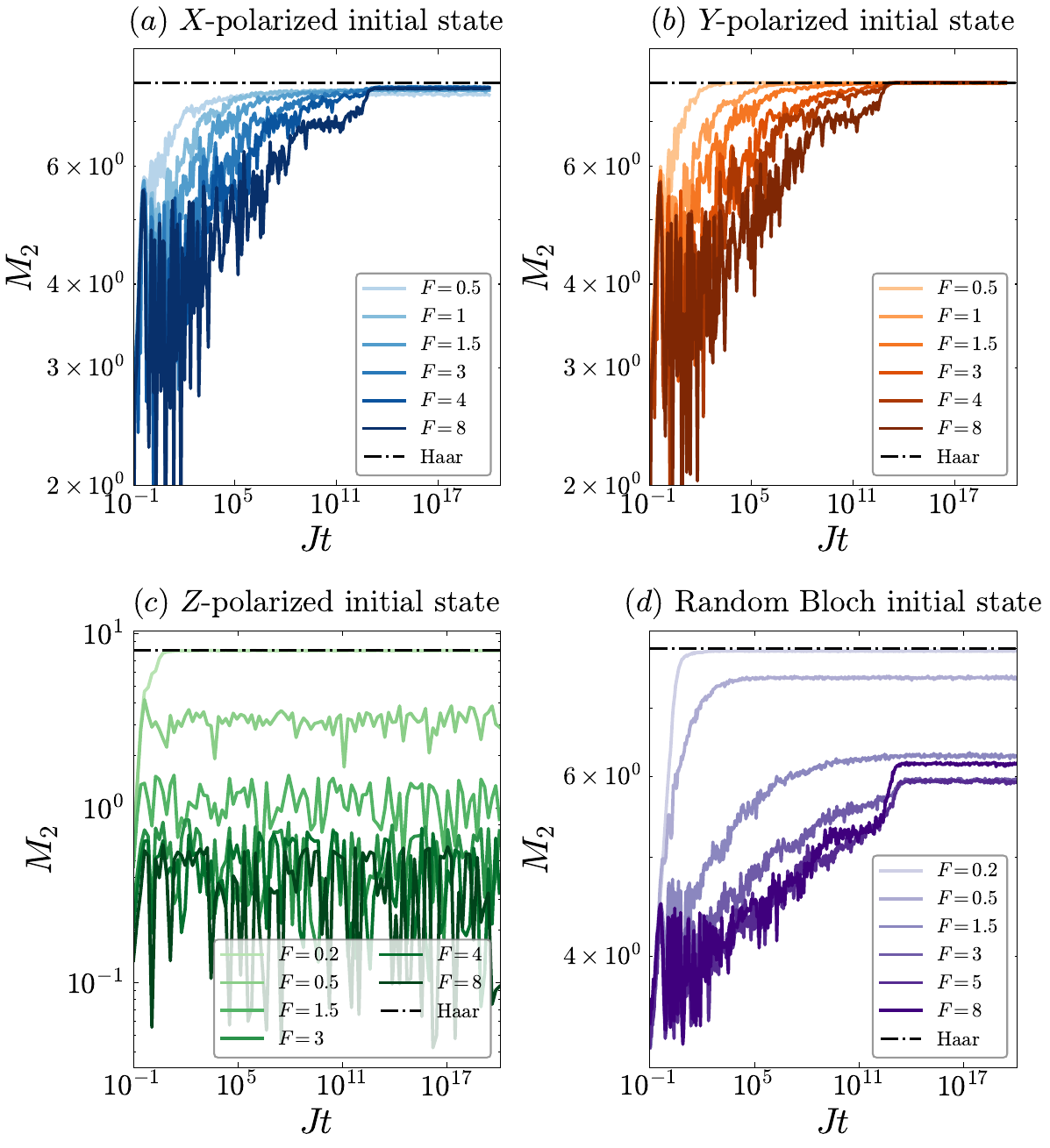}
\caption{\emph{Time evolution of the 2-SRE (magic).}
2-SRE $M_2(t)$ as a function of the dimensionless time $Jt$ for a fixed system size $L\!=\!10$ and four initial product states:
(a) $|\Psi_X\rangle\!=\!|+\rangle^{\otimes L}$,
(b) $|\Psi_Y\rangle\!=\!|+_y\rangle^{\otimes L}$,
(c) a $Z$-polarized computational-basis state,
and (d) random Bloch-sphere product states (ensemble-averaged; see Sec.~\ref{sec4}).
Colored curves correspond to different Stark gradients $F$ (see legends), with all other Hamiltonian parameters fixed as specified in Sec.~\ref{sec4}.
The black dash-dotted line indicates the Haar benchmark $M_2^{\mathrm{Haar}}$ in Eq.~\eqref{eq:M2_haar} for $L=10$.}
\label{fig1}
\end{figure}
\begin{figure*}[bt]
\hspace*{-0.98\textwidth}
\centering
\includegraphics[width=0.98\textwidth]{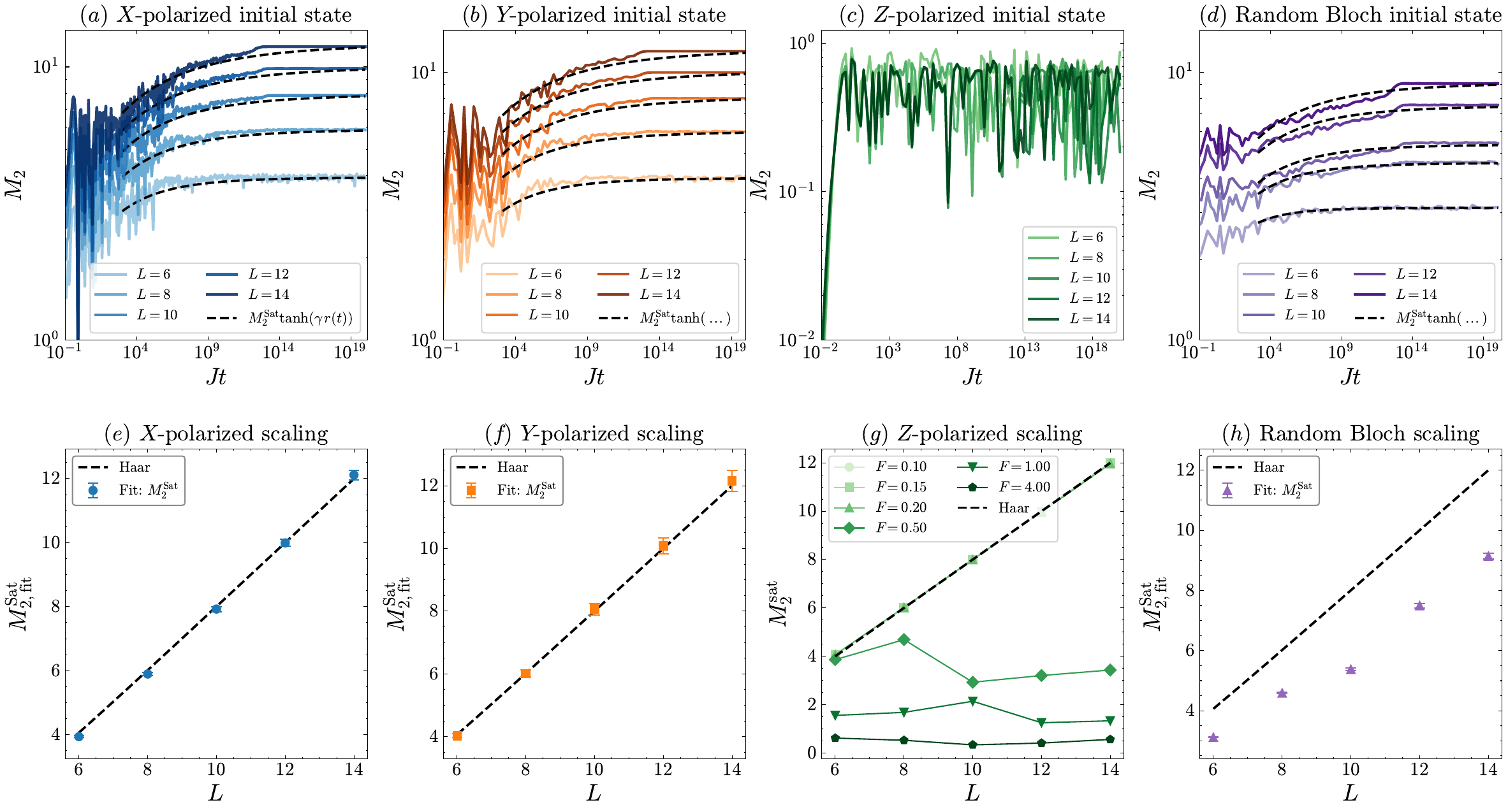}
\caption{\emph{Slow growth and finite-size scaling of the 2-SRE.}
(a--d) Time evolution of $M_2(t)$ for (a) $|\Psi_X\rangle$, (b) $|\Psi_Y\rangle$, (c) a $Z$-polarized computational-basis state, and (d) random Bloch-sphere product states (ensemble-averaged).
Curves correspond to system sizes $L=6,8,10,12,14$ (light to dark), at fixed Hamiltonian parameters (see Sec.~\ref{sec4}).
Black dashed curves in (a), (b), and (d) indicate fits to Eq.~\eqref{eq:log_growth} over the fitting window specified in Sec.~\ref{sec4}.
(e,f,h) Scaling of the fitted plateau values $M_{2,\mathrm{fit}}^{\mathrm{sat}}$ versus $L$ for the (e) $|\Psi_X\rangle$, (f) $|\Psi_Y\rangle$, and (h) random initial-state ensembles; the black dashed line shows $M_2^{\mathrm{Haar}}$ from Eq.~\eqref{eq:M2_haar}.
(g) Saturation value $M_2^{\mathrm{sat}}$ versus $L$ for several $F$ values (as indicated), illustrating a crossover from near-volume-law scaling at small $F$ to strongly suppressed, weakly size-dependent behavior at large $F$.}
\label{fig2}
\end{figure*}
\begin{figure}[bt]
\hspace*{-0.47\textwidth}
\centering
\includegraphics[width=0.47\textwidth]{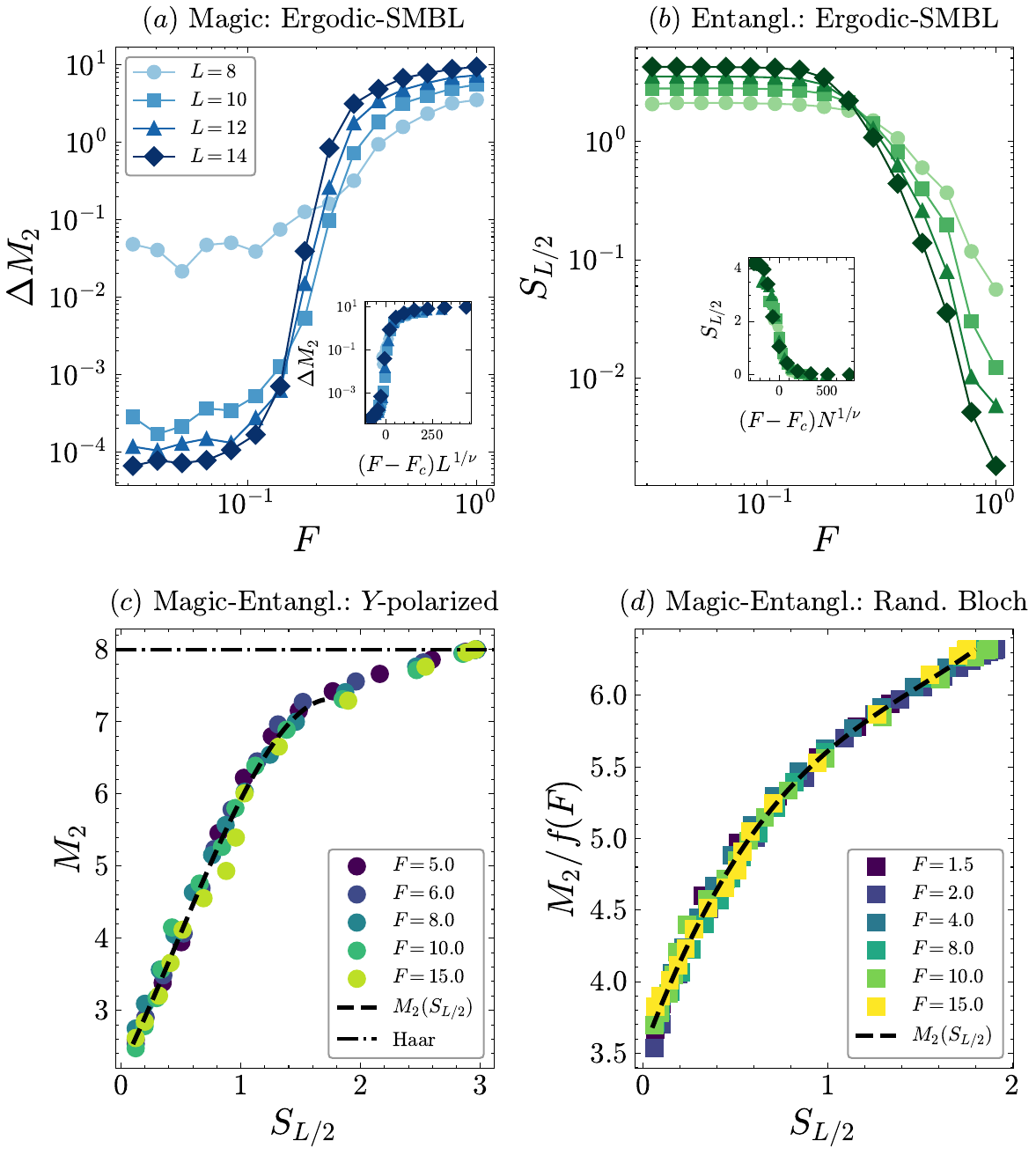}
\caption{\emph{Finite-size crossing of magic and entanglement.}
(a) Deviation from the Haar benchmark, $\Delta M_2=M_2^{\mathrm{Haar}}-M_2$, versus $F$ for $L=8,10,12,14$, where $M_2$ is taken from long-time values as specified in Sec.~\ref{sec4}.
Inset: finite-size collapse of $\Delta M_2$ versus $(F-F_c)L^{1/\nu}$ with $F_c\approx 0.19$ and $\nu\approx 0.53$.
(b) Half-chain entanglement entropy $S_{L/2}$ versus $F$ for the same sizes.
Inset: collapse of $S_{L/2}$ versus $(F-F_c)L^{1/\nu}$ with $F_c\approx 0.23$ and $\nu\approx 0.46$.
(c) Parametric plot of $M_2$ versus $S_{L/2}$ for the $|\Psi_Y\rangle$ initial state at fixed $L=10$ and several $F$ values; trajectories are obtained from time traces after Gaussian smoothing (kernel width specified in Sec.~\ref{sec4}).
The dashed curve is a fit guided by Eq.~\eqref{eq:log_growth}, and the dash-dotted line marks $M_2^{\mathrm{Haar}}$.
(d) Same as (c) for random Bloch-sphere product states, plotting the rescaled magic $M_2/f(F)$ with an $L$-independent factor $f(F)$; the dashed curve shows the corresponding fit.}
\label{fig3}
\end{figure}

This section examines how the Stark gradient $F$ controls the slow growth and saturation of quantum magic $M_2(t)$, its dependence on initial states and system size $L$, and its relation to entanglement dynamics. As a reference for complete randomization, we use the Haar benchmark $M_2^{\mathrm{Haar}}$ in Eq.~\eqref{eq:M2_haar}, and quantify the deviation by $\Delta M_2$ in Eq.~\eqref{eq:deltaM2}. 
All numerical realizations were carried out with \texttt{TensorCircuit-NG}~\cite{Zhang_2023} and \texttt{Qiskit}~\cite{qiskit}. Time evolution is implemented in the computational basis with controlled numerical precision, and the 2-SRE is evaluated using an efficient XOR fast Walsh--Hadamard-transform (XOR-FWHT)~\cite{xorfwht} based routine rather than explicit Pauli-string enumeration. For random Bloch-sphere product states, we average observables over an ensemble of independently sampled initial states. Full numerical details, including the XOR-FWHT evaluation of $M_2$, ensemble sampling, plateau extraction, and validation tests, are provided in Appendix~\ref{app:numerics}.

A first robust observation is that increasing $F$ systematically slows the generation of magic and, for certain initial states, also suppresses the long-time plateau. At fixed size $L\!=\!10$, both $|\Psi_X\rangle$ and $|\Psi_Y\rangle$ exhibit a rapid initial rise of $M_2(t)$ followed by a broad slow-growth regime and eventual saturation; the approach to the plateau becomes dramatically slower as $F$ increases [Fig.~\ref{fig1}(a,b)]. Random Bloch-sphere product states display the same qualitative slow-growth-and-saturation pattern [Fig.~\ref{fig1}(d)], but with a stronger $F$-dependence of the plateau: for larger tilts the saturation value can remain well below $M_2^{\mathrm{Haar}}$, corresponding to a sizable long-time $\Delta M_2$. In sharp contrast, the $Z$-polarized computational-basis initial state stays at very small $M_2(t)$ throughout the accessible time window. It shows pronounced temporal fluctuations without a clear monotonic slow-growth trend [Fig.~\ref{fig1}(c)], demonstrating strong initial-state selectivity in the production of non-Clifford resources under the same constrained dynamics.

The system-size dependence confirms that the slow growth and saturation are generic for a broad class of initial states. For $|\Psi_X\rangle$, $|\Psi_Y\rangle$, and random product ensembles, the time traces for $L=6,8,10,12,14$ consistently show slow growth over extended time windows followed by saturation, with the plateau increasing with $L$ [Fig.~\ref{fig2}(a,b,d)]. We fit these curves using Eq.~\eqref{eq:log_growth} (black dashed lines in Fig.~\ref{fig2}(a,b,d)) to extract plateau values used in the scaling analysis below. By comparison, the $Z$-polarized case remains dominated by low-valued, strongly fluctuating $M_2(t)$ across all studied sizes [Fig.~\ref{fig2}(c)], motivating a separate emphasis on its saturation behavior versus $F$.

The extracted long-time plateaus reveal distinct finite-size scaling behaviors across initial-state families. For $|\Psi_X\rangle$ and $|\Psi_Y\rangle$, the fitted plateaus $M_{2,\mathrm{fit}}^{\mathrm{sat}}$ grow approximately linearly with $L$ and closely track the Haar benchmark over the accessible sizes [Fig.~\ref{fig2}(e,f)], indicating near-Haar, volume-law-like magic at long times. For random Bloch product states, the saturation values also increase with $L$ but remain systematically below the Haar benchmark [Fig.~\ref{fig2}(h)], consistent with a persistent deviation from complete randomization. The $Z$-polarized case shows a particularly strong $F$-dependence: at small $F$ the saturation values increase with $L$ and approach the Haar benchmark, whereas at larger $F$ the $L$-dependence is strongly suppressed and the plateau tends toward a much smaller, weakly size-dependent value [Fig.~\ref{fig2}(g)].

We next characterize the ergodic-to-SMBL crossover using long-time observables. The deviation $\Delta M_2$ increases sharply with $F$ and exhibits a finite-size crossing across $L=8,10,12,14$ [Fig.~\ref{fig3}(a)]; the inset shows a collapse with $F_c\approx 0.19$ and $\nu\approx 0.53$. As an entanglement-based counterpart, the half-chain entanglement entropy $S_{L/2}$ decreases markedly as $F$ increases [Fig.~\ref{fig3}(b)], and its inset shows an analogous collapse with $F_c\approx 0.23$ and $\nu\approx 0.46$. Taken together, these results indicate a crossover from an ETH-like regime to constrained, Stark-localized dynamics, while highlighting that $\Delta M_2$ and $S_{L/2}$ provide complementary diagnostics of the same dynamical regime change.

Finally, we examine the relationship between magic and entanglement through parametric plots of time evolution in the $(S_{L/2},M_2)$ plane. For the $|\Psi_Y\rangle$ initial state at fixed $L=10$, the trajectories for different tilts broadly fall onto a single-valued curve that is well described by Eq.~\eqref{eq:log_growth} [Fig.~\ref{fig3}(c)]. For random Bloch product states, the raw trajectories show systematic offsets as $F$ is varied; after rescaling $M_2\to M_2/f(F)$ with an $L$-independent factor $f(F)$, the data collapse improves, and the common relation is again captured by Eq.~\eqref{eq:log_growth} [Fig.~\ref{fig3}(d)]. This indicates that, in the SMBL regime, the growth of entanglement and the generation of magic remain tightly correlated. In contrast, the initial-state structure and tilt strength modify the quantitative mapping in a way that a simple rescaling cannot fully account for.

In summary, stronger tilts generically slow down the growth of $M_2(t)$ and can reduce its long-time plateau; the saturation scaling is strongly initial-state dependent, with pronounced suppression for $Z$-polarized initial states at large $F$; and long-time behaviors of $\Delta M_2$ and $S_{L/2}$ consistently support a crossover from ETH-like to Stark-localized dynamics, alongside a clear functional relation between $M_2$ and $S_{L/2}$ in the SMBL regime.

\section{Experimental realization}\label{sec5}
\begin{figure*}[bt]
\hspace*{-0.98\textwidth}
\centering
\includegraphics[width=0.98\textwidth]{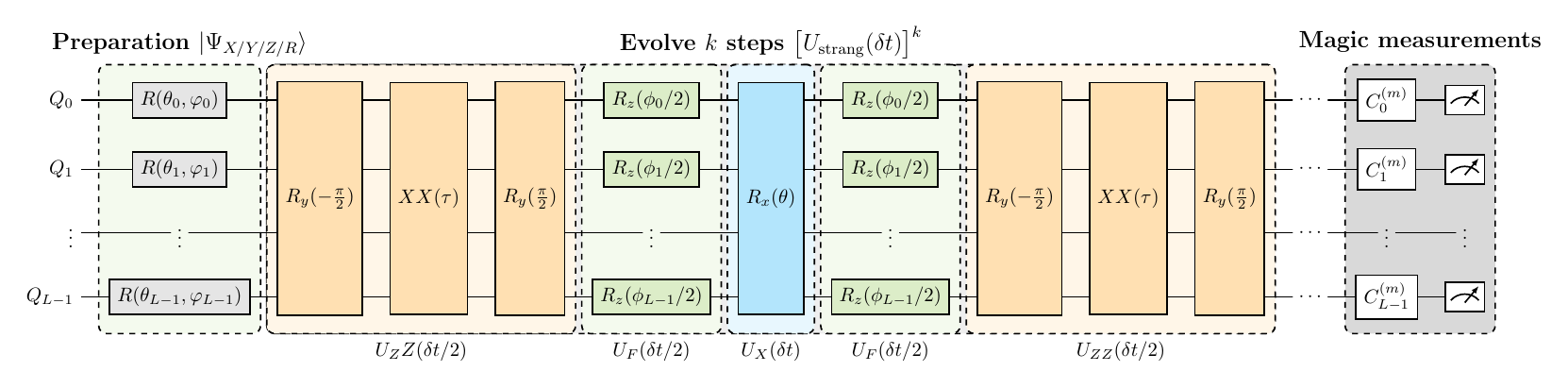}
\caption{\emph{Digital simulation and measurement protocol.}
Circuit schematic for digitally simulating the tilted transverse-field Ising model using a second-order Strang symmetric Trotter step repeated $k$ times, followed by local single-qubit Clifford randomized measurements. The same bitstring data are used to estimate the half-chain 2-R\'enyi entanglement entropy $S_2(A,t)$ and 2-SRE. See Appendix~\ref{app:exp_access} for gate compilation, unbiased estimators, and uncertainty analysis.}
\label{fig4}
\end{figure*}

We outline an experimentally viable protocol in a linear trapped-ion chain~\cite{exp1,exp2,Oliviero_2022} to digitally simulate transverse-field Ising dynamics in the presence of a purely linear Stark tilt and, from a single measurement data set, extract both the half-chain 2-R\'enyi entanglement entropy and the 2-SRE $M_2$. Calibration steps, estimator formulas, and error analysis are provided in Appendix~\ref{app:exp_access}.

We consider a chain of $L$ ions encoding $L$ qubits, with computational basis states $\ket{0}$ and $\ket{1}$ defined by state-dependent fluorescence detection. Ion-resolved readout implements a projective measurement in the $Z$ basis and returns a bitstring $s\!\in\!\{0,1\}^L$. Ions are indexed along the chain as $i\!=\!0,\ldots,L-1$, which fixes the spatial profile of the linear tilt. The target Hamiltonian is
\begin{align}
H=\sum_{i<j} J_{ij} Z_i Z_j \;+\; h \sum_{i=0}^{L-1} X_i \;+\; F \sum_{i=0}^{L-1} i\, Z_i,
\label{eq:H_target}
\end{align}
where $J_{ij}$ are effective long-range Ising couplings, $h$ is the transverse-field strength, and $F$ is the Stark-gradient parameter. We sample the dynamics at discrete times $t_k=k\,\delta t$ for integers $k\ge 0$, with a digital step size $\delta t$.

A calibration stage precedes the dynamical experiment to map experimental controls onto $(J_{ij},h,F)$. The transverse field $h$ is determined from single-qubit Rabi oscillations, which fix the pulse-area-to-rotation-angle conversion and thereby the transverse-field step angle $\theta=2h\,\delta t$. The coupling matrix $J_{ij}$ can be obtained by enabling the entangling interaction for controlled durations and extracting accumulated two-body phases and/or correlation signals. The linear tilt $F$ is calibrated through ion-resolved Ramsey phase accumulation: the measured $Z$-frequency shifts scale linearly with the ion index $i$, and any deviation from strict linearity is treated as a systematic uncertainty (Appendix~\ref{app:exp_access}).

To implement time evolution, we use a second-order symmetric (Strang) decomposition. A native long-range two-body interaction is available in trapped ions and, together with global basis rotations, yields an effective $ZZ$ evolution realizing the $\sum_{i<j}J_{ij}Z_iZ_j$ term. The transverse-field contribution is implemented as a global single-qubit rotation $U_X(\delta t)=\prod_i R_x^{(i)}(\theta)$. The Stark tilt yields a diagonal evolution equivalent to a product of local $Z$ rotations with ion-dependent phases $\phi_i=2Fi\,\delta t$, which can be implemented via ion-resolved AC-Stark shifts or, equivalently, tracked as phase-frame updates. Grouping the two-body $ZZ$ evolution and the tilt into a single diagonal block $U_{\mathrm{diag}}(\tau)$, we define one symmetric digital step as
\begin{align}
U_{\mathrm{step}}(\delta t)&=U_{\mathrm{diag}}(\delta t/2)\,U_X(\delta t)\,U_{\mathrm{diag}}(\delta t/2),\\
U(t_k)&\approx \bigl[U_{\mathrm{step}}(\delta t)\bigr]^k,
\label{eq:step}
\end{align}
with $t_k=k\,\delta t$. The gate ordering, timing windows, and dominant error sources (Trotter error, coherent control errors, and decoherence) are discussed in Appendix~\ref{app:exp_access}.

We probe both structured and typical dynamical regimes by preparing several families of product initial states, including a fully $Z$-polarized state, the transverse-polarized states $|\Psi_X\rangle=|+\rangle^{\otimes L}$ and $|\Psi_Y\rangle=|+_y\rangle^{\otimes L}$, and an ensemble of random product states generated by independent single-qubit Bloch-sphere sampling. For the random ensemble, we repeat the full protocol for each sampled initial state and report observables averaged over the ensemble to suppress finite-size fluctuations (Appendix~\ref{app:exp_access}).

At each time point $t_k$, we extract both entanglement and nonstabilizerness using a local randomized measurement protocol based on single-qubit Clifford operations followed by computational-basis readout. Concretely, we sample $N_U$ local Clifford settings $C^{(m)}=\bigotimes_{i=0}^{L-1} c_i^{(m)}$ with $c_i^{(m)}\in\mathcal{C}_1$ and $m=1,\ldots,N_U$, apply $C^{(m)}$ to the evolved state $\rho(t_k)$, and measure all qubits in the $Z$ basis $N_M$ times to obtain bitstrings $\{s^{(m)}_\alpha\}_{\alpha=1}^{N_M}$.

The half-chain 2-R\'enyi entanglement entropy is $S_2(A,t)=-\log_2 \Tr[\rho_A(t)^2]$, where $\rho_A(t)=\Tr_{A^c}\rho(t)$ and $A$ denotes the left half of the chain. Using the same bitstring data, we estimate the subsystem purity $\Tr[\rho_A^2]$ via collision statistics built from Hamming-weight kernels and set $S_2(A,t_k)=-\log_2 P_A(t_k)$. From the same data, we also estimate the global purity $\Tr[\rho(t_k)^2]$, which serves both as a diagnostic of decoherence and as a mixedness correction for the magic estimator. Explicit unbiased estimators are given in Appendix~\ref{app:exp_access}.

To quantify magic, we use the mixed-state-consistent definition of the 2-SRE in terms of the Pauli fourth moment. Let $\mathcal{P}_L$ denote the $L$-qubit Pauli operators up to phase and define $W(t_k)=D^{-2}\sum_{P\in\mathcal{P}_L}\langle P\rangle_{t_k}^{4}$, where $\langle P\rangle_{t_k}=\Tr[P\,\rho(t_k)]$ and $D=2^L$. We then evaluate
\begin{align}
M_2(t_k)=-\log_2\!\left[\frac{D\,W(t_k)}{\Tr\!\bigl(\rho(t_k)^2\bigr)}\right],
\label{eq:M2_def}
\end{align}
which reduces to $M_2\simeq -\log_2(DW)$ when $\rho(t_k)$ remains close to pure. In practice, $W(t_k)$ is estimated from the same randomized-measurement data using a four-bitstring Hamming-weight kernel, which requires $N_M\ge 4$, and combined with the estimated global purity to correct for experimental mixedness (Appendix~\ref{app:exp_access}). The total number of single-shot measurements per time point is $N_{\mathrm{tot}}=N_U N_M$. Increasing $N_U$ primarily reduces fluctuations from finite unitary sampling, whereas increasing $N_M$ primarily reduces within-setting shot noise. Statistical uncertainties are obtained via bootstrap resampling over the Clifford-setting index $m$, which preserves the estimator structure while providing robust error bars. A circuit-level overview of the protocol is shown in Fig.~\ref{fig4}.

\section{Conclusion and outlook}\label{sec6}
In this work, we studied the dynamical generation of nonstabilizerness in a clean system exhibiting Stark many-body localization. Using the second stabilizer R\'enyi entropy $M_2$ and standard entanglement diagnostics for a tilted transverse-field Ising chain, we find that in the strong-tilt SMBL regime, $M_2(t)$ remains nonzero and exhibits a robust, slow growth before saturating to a finite-size plateau, with pronounced initial-state selectivity. Long-time magic and entanglement consistently indicate a crossover from an ETH-like regime to constrained localized dynamics as the tilt increases, establishing nonstabilizerness as a practical complexity probe for disorder-free ergodicity breaking. In the strong-tilt limit, a Schrieffer--Wolff effective diagonal description yields emergent couplings that are factorially suppressed along the Wannier--Stark ladder, providing a microscopic origin for the slow dephasing front and the resulting slow growth of $M_2(t)$. Finally, we outlined a trapped-ion digital-simulation protocol in which both the half-chain R\'enyi-2 entanglement entropy and $M_2$ can be extracted from the same local randomized-measurement data with a mixedness correction.

An immediate direction is to extend these ideas to other constrained and monitored settings, and to explore possible connections between nonstabilizerness dynamics and the quantum Mpemba effect~\cite{LiuZhangYinZhang2024PRL,QME1,QME3,sxz4,Yu_2025,xhek_QME1,xhek_QME2,xhek_QME3}.

\begin{acknowledgements}
J.-X.\ Zhong was supported by the National Natural Science Foundation of China (Grant Nos.\ 12374046 and 11874316), the Shanghai Science and Technology Innovation Action Plan (Grant No.\ 24LZ1400800), the National Basic Research Program of China (Grant No.\ 2015CB921103), and the Program for Changjiang Scholars and Innovative Research Teams in Universities (Grant No.\ IRT13093). 
H.-Z.\ Li is supported by a China Scholarship Council Scholarship.
 
\end{acknowledgements}

%%%%%%%%%%%%%%%%% APPENDIX %%%%%%%%%%%%%%%%%%%
\bigskip
\bibliography{refs}

@article{ClaeysLamacraft2022Quantum,
  author = {Claeys, P. W. and Lamacraft, A.},
  title = {Quantum many-body scars and space-time crystalline order from magnon condensation},
  journal = {Phys. Rev. Lett.},
  year = {2022},
  volume = {129},
  pages = {100601},
  doi = {10.1103/PhysRevLett.129.100601}
}

@article{Feynman1982Simulating,
  author  = {Feynman, Richard P.},
  title   = {Simulating Physics with Computers},
  journal = {International Journal of Theoretical Physics},
  year    = {1982},
  volume  = {21},
  number  = {6/7},
  pages   = {467--488},
  doi     = {10.1007/BF02650179}
}

@article{Lloyd1996UniversalSimulators,
  author  = {Lloyd, Seth},
  title   = {Universal Quantum Simulators},
  journal = {Science},
  year    = {1996},
  volume  = {273},
  number  = {5278},
  pages   = {1073--1078},
  doi     = {10.1126/science.273.5278.1073}
}

@article{Quantinuum2025,
  title = {Experimental Demonstration of High-Fidelity Logical Magic States from Code Switching},
  author = {Daguerre, Lucas and Blume-Kohout, Robin and Brown, Natalie C. and Hayes, David and Kim, Isaac H.},
  journal = {Phys. Rev. X},
  volume = {15},
  issue = {4},
  pages = {041008},
  numpages = {11},
  year = {2025},
  month = {Oct},
  publisher = {American Physical Society},
  doi = {10.1103/dck4-x9c2},
  url = {https://link.aps.org/doi/10.1103/dck4-x9c2}
}

@article{Aasen2025,
      title={Roadmap to fault tolerant quantum computation using topological qubit arrays}, 
      author={David Aasen and Morteza Aghaee and Zulfi Alam and Mariusz Andrzejczuk and Andrey Antipov and Mikhail Astafev and Lukas Avilovas and Amin Barzegar and Bela Bauer and Jonathan Becker and Juan M. Bello-Rivas and Umesh Bhaskar and Alex Bocharov and Srini Boddapati and David Bohn and others},
      year={2025},
      journal={arXiv:2502.12252},
      archivePrefix={arXiv},
      primaryClass={quant-ph},
      url={https://arxiv.org/abs/2502.12252}, 
}

@article{Peham2025,
  title = {Automated Synthesis of Fault-Tolerant State Preparation Circuits for Quantum Error-Correction Codes},
  author = {Peham, Tom and Schmid, Ludwig and Berent, Lucas and M\"uller, Markus and Wille, Robert},
  journal = {PRX Quantum},
  volume = {6},
  issue = {2},
  pages = {020330},
  numpages = {32},
  year = {2025},
  month = {May},
  publisher = {American Physical Society},
  doi = {10.1103/PRXQuantum.6.020330},
  url = {https://link.aps.org/doi/10.1103/PRXQuantum.6.020330}
}

@article{Shor1997Algorithms,
  author  = {Shor, Peter W.},
  title   = {Polynomial-Time Algorithms for Prime Factorization and Discrete Logarithms on a Quantum Computer},
  journal = {SIAM Journal on Computing},
  year    = {1997},
  volume  = {26},
  number  = {5},
  pages   = {1484--1509},
  doi     = {10.1137/S0097539795293172}
}

@article{Grover1997Search,
  author  = {Grover, Lov K.},
  title   = {Quantum Mechanics Helps in Searching for a Needle in a Haystack},
  journal = {Physical Review Letters},
  year    = {1997},
  volume  = {79},
  number  = {2},
  pages   = {325--328},
  doi     = {10.1103/PhysRevLett.79.325}
}

@article{BernsteinVazirani1997QCT,
  author  = {Bernstein, Ethan and Vazirani, Umesh},
  title   = {Quantum Complexity Theory},
  journal = {SIAM Journal on Computing},
  year    = {1997},
  volume  = {26},
  number  = {5},
  pages   = {1411--1473},
  doi     = {10.1137/S0097539796300921}
}

@inproceedings{AaronsonArkhipov2011LinearOptics,
  author    = {Aaronson, Scott and Arkhipov, Alex},
  title     = {The Computational Complexity of Linear Optics},
  booktitle = {Proceedings of the 43rd ACM Symposium on Theory of Computing (STOC)},
  year      = {2011},
  pages     = {333--342},
  doi       = {10.1145/1993636.1993682}
}

@article{Preskill2018NISQ,
  author  = {Preskill, John},
  title   = {Quantum Computing in the {NISQ} Era and Beyond},
  journal = {Quantum},
  year    = {2018},
  volume  = {2},
  pages   = {79},
  doi     = {10.22331/q-2018-08-06-79}
}

@article{Zhong2020PhotonicAdvantage,
  author  = {Zhong, Han-Sen and others},
  title   = {Quantum Computational Advantage Using Photons},
  journal = {Science},
  year    = {2020},
  volume  = {370},
  number  = {6523},
  pages   = {1460--1463},
  doi     = {10.1126/science.abe8770}
}

@article{Kretschmer2025,
Author = {William Kretschmer and Sabee Grewal and Matthew DeCross and Justin A. Gerber and Kevin Gilmore and Dan Gresh and Nicholas Hunter-Jones and Karl Mayer and Brian Neyenhuis and David Hayes and Scott Aaronson},
Title = {Demonstrating an unconditional separation between quantum and classical information resources},
Year = {2025},
journal={arXiv:2509.07255},
archivePrefix={arXiv},
primaryClass={quant-ph},
url={https://arxiv.org/abs/2509.07255}, 
}

@article{Niroula2024,
  title = {Phase transition in magic with random quantum circuits},
  volume = {20},
  ISSN = {1745-2481},
  url = {http://dx.doi.org/10.1038/s41567-024-02637-3},
  DOI = {10.1038/s41567-024-02637-3},
  number = {11},
  journal = {Nature Physics},
  publisher = {Springer Science and Business Media LLC},
  author = {Niroula,  Pradeep and White,  Christopher David and Wang,  Qingfeng and Johri,  Sonika and Zhu,  Daiwei and Monroe,  Christopher and Noel,  Crystal and Gullans,  Michael J.},
  year = {2024},
  month = sep,
  pages = {1786–1792}
}

@article{Nahum2018,
  title = {Operator Spreading in Random Unitary Circuits},
  author = {Nahum, Adam and Vijay, Sagar and Haah, Jeongwan},
  journal = {Phys. Rev. X},
  volume = {8},
  issue = {2},
  pages = {021014},
  numpages = {30},
  year = {2018},
  month = {Apr},
  publisher = {American Physical Society},
  doi = {10.1103/PhysRevX.8.021014},
  url = {https://link.aps.org/doi/10.1103/PhysRevX.8.021014}
}

@article{tirrito2025anticoncentrationnonstabilizernessspreadingergodic,
      title={Anticoncentration and nonstabilizerness spreading under ergodic quantum dynamics}, 
      author={Emanuele Tirrito and Xhek Turkeshi and Piotr Sierant},
      year={2025},
      journal={arXiv:2412.10229},
      archivePrefix={arXiv},
      primaryClass={quant-ph},
      url={https://arxiv.org/abs/2412.10229}, 
}

@article{zhang2025stabilizerrenyientropytransition,
      title={Stabilizer R\'enyi Entropy and its Transition in the Coupled Sachdev-Ye-Kitaev Model}, 
      author={Pengfei Zhang and Shuyan Zhou and Ning Sun},
      year={2025},
      journal={arXiv:2509.17417},
      archivePrefix={arXiv},
      primaryClass={quant-ph},
      url={https://arxiv.org/abs/2509.17417}, 
}

@article{qian2025quantum,
  title = {Quantum nonlocal nonstabilizerness},
  author = {Qian, Dongheng and Wang, Jing},
  journal = {Phys. Rev. A},
  volume = {111},
  issue = {5},
  pages = {052443},
  numpages = {9},
  year = {2025},
  month = {May},
  publisher = {American Physical Society},
  doi = {10.1103/PhysRevA.111.052443},
  url = {https://link.aps.org/doi/10.1103/PhysRevA.111.052443}
}

@article{Tarabunga_2024,
   title={Magic in generalized Rokhsar-Kivelson wavefunctions},
   volume={8},
   ISSN={2521-327X},
   url={http://dx.doi.org/10.22331/q-2024-05-14-1347},
   DOI={10.22331/q-2024-05-14-1347},
   journal={Quantum},
   publisher={Verein zur Forderung des Open Access Publizierens in den Quantenwissenschaften},
   author={Tarabunga, Poetri Sonya and Castelnovo, Claudio},
   year={2024},
   month=may, pages={1347} }

@article{qiant25,
  title = {Augmenting a finite-temperature tensor network with Clifford circuits},
  author = {Qian, Xiangjian and Huang, Jiale and Qin, Mingpu},
  journal = {Phys. Rev. B},
  volume = {112},
  issue = {11},
  pages = {115150},
  numpages = {8},
  year = {2025},
  month = {Sep},
  publisher = {American Physical Society},
  doi = {10.1103/gljy-1ykf},
  url = {https://link.aps.org/doi/10.1103/gljy-1ykf}
}

@article{qian2025,
  title = {Clifford Circuits Augmented Time-Dependent Variational Principle},
  author = {Qian, Xiangjian and Huang, Jiale and Qin, Mingpu},
  journal = {Phys. Rev. Lett.},
  volume = {134},
  issue = {15},
  pages = {150404},
  numpages = {6},
  year = {2025},
  month = {Apr},
  publisher = {American Physical Society},
  doi = {10.1103/PhysRevLett.134.150404},
  url = {https://link.aps.org/doi/10.1103/PhysRevLett.134.150404}
}

@article{hang2025,
  title = {Nonstabilizerness entanglement entropy: A measure of hardness in the classical simulation of quantum many-body systems with tensor network states},
  author = {Huang, Jiale and Qian, Xiangjian and Qin, Mingpu},
  journal = {Phys. Rev. A},
  volume = {112},
  issue = {1},
  pages = {012425},
  numpages = {11},
  year = {2025},
  month = {Jul},
  publisher = {American Physical Society},
  doi = {10.1103/gxdn-zwrw},
  url = {https://link.aps.org/doi/10.1103/gxdn-zwrw}
}

@article{qiant2024a,
  title = {Augmenting Density Matrix Renormalization Group with Clifford Circuits},
  author = {Qian, Xiangjian and Huang, Jiale and Qin, Mingpu},
  journal = {Phys. Rev. Lett.},
  volume = {133},
  issue = {19},
  pages = {190402},
  numpages = {6},
  year = {2024},
  month = {Nov},
  publisher = {American Physical Society},
  doi = {10.1103/PhysRevLett.133.190402},
  url = {https://link.aps.org/doi/10.1103/PhysRevLett.133.190402}
}

@Article{frau2025,
	title={{Stabilizer disentangling of conformal field theories}},
	author={Martina Frau and Poetri Sonya Tarabunga and Mario Collura and Emanuele Tirrito and Marcello Dalmonte},
	journal={SciPost Phys.},
	volume={18},
	pages={165},
	year={2025},
	publisher={SciPost},
	doi={10.21468/SciPostPhys.18.5.165},
	url={https://scipost.org/10.21468/SciPostPhys.18.5.165},
}

@article{fan2025disentangling,
  title = {Disentangling critical quantum spin chains with Clifford circuits},
  author = {Fan, Chaohui and Qian, Xiangjian and Zhang, Hua-Chen and Huang, Rui-Zhen and Qin, Mingpu and Xiang, Tao},
  journal = {Phys. Rev. B},
  volume = {111},
  issue = {8},
  pages = {085121},
  numpages = {8},
  year = {2025},
  month = {Feb},
  publisher = {American Physical Society},
  doi = {10.1103/PhysRevB.111.085121},
  url = {https://link.aps.org/doi/10.1103/PhysRevB.111.085121}
}

@article{huang2024cliffordcircuitsaugmentedmatrix,
      title={Clifford circuits Augmented Matrix Product States for fermion systems}, 
      author={Jiale Huang and Xiangjian Qian and Mingpu Qin},
      year={2024},
      journal={arXiv:2501.00413},
      archivePrefix={arXiv},
      primaryClass={cond-mat.str-el},
      url={https://arxiv.org/abs/2501.00413}, 
}

@article{korbany2025longrangenonstabilizernessphasesmatter,
      title={Long-range nonstabilizerness and phases of matter}, 
      author={David Aram Korbany and Michael J. Gullans and Lorenzo Piroli},
      year={2025},
      journal={arXiv:2502.19504},
      archivePrefix={arXiv},
      primaryClass={quant-ph},
      url={https://arxiv.org/abs/2502.19504}, 
}

@article{ding2025evaluating,
  title = {Evaluating Many-Body Stabilizer R\'enyi Entropy by Sampling Reduced Pauli Strings: Singularities, Volume Law, and Nonlocal Magic},
  author = {Ding, Yi-Ming and Wang, Zhe and Yan, Zheng},
  journal = {PRX Quantum},
  volume = {6},
  issue = {3},
  pages = {030328},
  numpages = {15},
  year = {2025},
  month = {Aug},
  publisher = {American Physical Society},
  doi = {10.1103/pyzr-jmvw},
  url = {https://link.aps.org/doi/10.1103/pyzr-jmvw}
}

@article{tarabunga2025efficientmutualmagicmagic,
      title={Efficient mutual magic and magic capacity with matrix product states}, 
      author={Poetri Sonya Tarabunga and Tobias Haug},
      year={2025},
      journal={arXiv:2504.07230},
      archivePrefix={arXiv},
      primaryClass={quant-ph},
      url={https://arxiv.org/abs/2504.07230}, 
}

@article{cao2025gravitationalbackreactionmagical,
      title={Gravitational back-reaction is magical}, 
      author={ChunJun Cao and Gong Cheng and Alioscia Hamma and Lorenzo Leone and William Munizzi and Savatore F. E. Oliviero},
      year={2025},
      journal={arXiv:2403.07056},
      archivePrefix={arXiv},
      primaryClass={hep-th},
      url={https://arxiv.org/abs/2403.07056}, 
}

@article{moca2025nonstabilizernessgenerationmultiparticlequantum,
      title={Non-stabilizerness generation in a multi-particle quantum walk}, 
      author={Cătălin Paşcu Moca and Doru Sticlet and Balázs Dóra and Angelo Valli and Dominik Szombathy and Gergely Zaránd},
      year={2025},
      journal={arXiv:2504.19750},
      archivePrefix={arXiv},
      primaryClass={quant-ph},
      url={https://arxiv.org/abs/2504.19750}, 
}

@article{Haug_2023_1,
   title={{Quantifying nonstabilizerness of matrix product states}},
   volume={107},
   ISSN={2469-9969},
   url={http://dx.doi.org/10.1103/PhysRevB.107.035148},
   DOI={10.1103/physrevb.107.035148},
   number={3},
   journal={Physical Review B},
pages={035148},
   publisher={American Physical Society (APS)},
   author={Haug, Tobias and Piroli, Lorenzo},
   year={2023},
   month={1}
}

@article{tarabunga2023manybody,
  title = {Many-Body Magic Via Pauli-Markov Chains---From Criticality to Gauge Theories},
  author = {Tarabunga, Poetri Sonya and Tirrito, Emanuele and Chanda, Titas and Dalmonte, Marcello},
  journal = {PRX Quantum},
  volume = {4},
  issue = {4},
  pages = {040317},
  numpages = {19},
  year = {2023},
  month = {Oct},
  publisher = {American Physical Society},
  doi = {10.1103/PRXQuantum.4.040317},
  url = {https://link.aps.org/doi/10.1103/PRXQuantum.4.040317}
}

@article{tarabunga2024nonstabilizerness,
  title = {Nonstabilizerness via Matrix Product States in the Pauli Basis},
  author = {Tarabunga, Poetri Sonya and Tirrito, Emanuele and Ba\~nuls, Mari Carmen and Dalmonte, Marcello},
  journal = {Phys. Rev. Lett.},
  volume = {133},
  issue = {1},
  pages = {010601},
  numpages = {7},
  year = {2024},
  month = {Jul},
  publisher = {American Physical Society},
  doi = {10.1103/PhysRevLett.133.010601},
  url = {https://link.aps.org/doi/10.1103/PhysRevLett.133.010601}
}

@article{tirrito2024quantifying,
  title = {Quantifying nonstabilizerness through entanglement spectrum flatness},
  author = {Tirrito, Emanuele and Tarabunga, Poetri Sonya and Lami, Gugliemo and Chanda, Titas and Leone, Lorenzo and Oliviero, Salvatore F. E. and Dalmonte, Marcello and Collura, Mario and Hamma, Alioscia},
  journal = {Phys. Rev. A},
  volume = {109},
  issue = {4},
  pages = {L040401},
  numpages = {6},
  year = {2024},
  month = {Apr},
  publisher = {American Physical Society},
  doi = {10.1103/PhysRevA.109.L040401},
  url = {https://link.aps.org/doi/10.1103/PhysRevA.109.L040401}
}

@article{Lami_2023_2,
  title = {{Nonstabilizerness via Perfect Pauli Sampling of Matrix Product States}},
  author = {Lami, Guglielmo and Collura, Mario},
  journal = {Phys. Rev. Lett.},
  volume = {131},
  issue = {18},
  pages = {180401},
  numpages = {6},
  year = {2023},
  month = {Oct},
  publisher = {American Physical Society},
  doi = {10.1103/PhysRevLett.131.180401},
  url = {https://link.aps.org/doi/10.1103/PhysRevLett.131.180401}
}

@article{Lami_2024,
  title = {{Unveiling the Stabilizer Group of a Matrix Product State}},
  author = {Lami, Guglielmo and Collura, Mario},
  journal = {Phys. Rev. Lett.},
  volume = {133},
  issue = {1},
  pages = {010602},
  numpages = {6},
  year = {2024},
  month = {Jul},
  publisher = {American Physical Society},
  doi = {10.1103/PhysRevLett.133.010602},
  url = {https://link.aps.org/doi/10.1103/PhysRevLett.133.010602}
}

@article{Leone_2024,
  title = {Stabilizer entropies are monotones for magic-state resource theory},
  author = {Leone, Lorenzo and Bittel, Lennart},
  journal = {Phys. Rev. A},
  volume = {110},
  issue = {4},
  pages = {L040403},
  numpages = {6},
  year = {2024},
  month = {Oct},
  publisher = {American Physical Society},
  doi = {10.1103/PhysRevA.110.L040403},
  url = {https://link.aps.org/doi/10.1103/PhysRevA.110.L040403}
}

@article{Turkeshi_2023,
   title={{Measuring nonstabilizerness via multifractal flatness}},
   volume={108},
   ISSN={2469-9934},
   url={http://dx.doi.org/10.1103/PhysRevA.108.042408},
   DOI={10.1103/physreva.108.042408},
   number={4},
   journal={Physical Review A},
   publisher={American Physical Society (APS)},
   author={Turkeshi, Xhek and Schirò, Marco and Sierant, Piotr},
   year={2023},
pages={042408},
   month=oct
}

@article{Haug2025probingquantum,
  doi = {10.22331/q-2025-07-21-1801},
  url = {https://doi.org/10.22331/q-2025-07-21-1801},
  title = {Probing quantum complexity via universal saturation of stabilizer entropies},
  author = {Haug, Tobias and Aolita, Leandro and Kim, M.S.},
  journal = {{Quantum}},
  issn = {2521-327X},
  publisher = {{Verein zur F{\"{o}}rderung des Open Access Publizierens in den Quantenwissenschaften}},
  volume = {9},
  pages = {1801},
  month = jul,
  year = {2025}
}

@article{jasser2025stabilizerentropyentanglementcomplexity,
      title={Stabilizer Entropy and entanglement complexity in the Sachdev-Ye-Kitaev model}, 
      author={Barbara Jasser and Jovan Odavić and Alioscia Hamma},
      year={2025},
      journal={arXiv:2502.03093},
      archivePrefix={arXiv},
      primaryClass={quant-ph},
      url={https://arxiv.org/abs/2502.03093}, 
}

@article{viscardi2025interplayentanglementstructuresstabilizer,
      title={Interplay of entanglement structures and stabilizer entropy in spin models}, 
      author={Michele Viscardi and Marcello Dalmonte and Alioscia Hamma and Emanuele Tirrito},
      year={2025},
      journal={arXiv:2503.08620},
      archivePrefix={arXiv},
      primaryClass={quant-ph},
      url={https://arxiv.org/abs/2503.08620}, 
}

@article{turkeshi2025pauli,
  title = {Pauli spectrum and nonstabilizerness of typical quantum many-body states},
  author = {Turkeshi, Xhek and Dymarsky, Anatoly and Sierant, Piotr},
  journal = {Phys. Rev. B},
  volume = {111},
  issue = {5},
  pages = {054301},
  numpages = {12},
  year = {2025},
  month = {Feb},
  publisher = {American Physical Society},
  doi = {10.1103/PhysRevB.111.054301},
  url = {https://link.aps.org/doi/10.1103/PhysRevB.111.054301}
}

@article{Iannotti2025entanglement,
  doi = {10.22331/q-2025-07-21-1797},
  url = {https://doi.org/10.22331/q-2025-07-21-1797},
  title = {Entanglement and {S}tabilizer entropies of random bipartite pure quantum states},
  author = {Iannotti, Daniele and Esposito, Gianluca and Campos Venuti, Lorenzo and Hamma, Alioscia},
  journal = {{Quantum}},
  issn = {2521-327X},
  publisher = {{Verein zur F{\"{o}}rderung des Open Access Publizierens in den Quantenwissenschaften}},
  volume = {9},
  pages = {1797},
  month = jul,
  year = {2025}
}

@article{cusumano2025nonstabilizernessviolationschshinequalities,
      title={Non-stabilizerness and violations of CHSH inequalities}, 
      author={Stefano Cusumano and Lorenzo Campos Venuti and Simone Cepollaro and Gianluca Esposito and Daniele Iannotti and Barbara Jasser and Jovan Odavi\' c and Michele Viscardi and Alioscia Hamma},
      year={2025},
      journal={arXiv:2504.03351},
      archivePrefix={arXiv},
      primaryClass={quant-ph},
      url={https://arxiv.org/abs/2504.03351}, 
}

@article{tarabunga2024magictransitionmeasurementonlycircuits,
      title={Magic transition in measurement-only circuits}, 
      author={Poetri Sonya Tarabunga and Emanuele Tirrito},
      year={2024},
      journal={arXiv:2407.15939},
      archivePrefix={arXiv},
      primaryClass={quant-ph},
      url={https://arxiv.org/abs/2407.15939}, 
}

@article{tirrito2025magicphasetransitionsmonitored,
      title={Magic phase transitions in monitored gaussian fermions}, 
      author={Emanuele Tirrito and Luca Lumia and Alessio Paviglianiti and Guglielmo Lami and Alessandro Silva and Xhek Turkeshi and Mario Collura},
      year={2025},
      journal={arXiv:2507.07179},
      archivePrefix={arXiv},
      primaryClass={quant-ph},
      url={https://arxiv.org/abs/2507.07179}, 
}

@article{wang2025magictransitionmonitoredfree,
      title={Magic transition in monitored free fermion dynamics}, 
      author={Cheng Wang and Zhi-Cheng Yang and Tianci Zhou and Xiao Chen},
      year={2025},
      journal={arXiv:2507.10688},
      archivePrefix={arXiv},
      primaryClass={quant-ph},
      url={https://arxiv.org/abs/2507.10688}, 
}

@article{santra2025complexitytransitionschaoticquantum,
      title={Complexity transitions in chaotic quantum systems}, 
      author={Gopal Chandra Santra and Alex Windey and Soumik Bandyopadhyay and Andrea Legramandi and Philipp Hauke},
      year={2025},
      journal={arXiv:2505.09707},
      archivePrefix={arXiv},
      primaryClass={quant-ph},
      url={https://arxiv.org/abs/2505.09707}, 
}

@article{bittel2025operationalinterpretationstabilizerentropy,
      title={Operational interpretation of the Stabilizer Entropy}, 
      author={Lennart Bittel and Lorenzo Leone},
      year={2025},
      journal={arXiv:2507.22883},
      archivePrefix={arXiv},
      primaryClass={quant-ph},
      url={https://arxiv.org/abs/2507.22883}, 
}

@article{varikuti2025impactcliffordoperationsnonstabilizing,
      title={Impact of Clifford operations on non-stabilizing power and quantum chaos}, 
      author={Naga Dileep Varikuti and Soumik Bandyopadhyay and Philipp Hauke},
      year={2025},
      journal={arXiv:2505.14793},
      archivePrefix={arXiv},
      primaryClass={quant-ph},
      url={https://arxiv.org/abs/2505.14793}, 
}

@article{dowling2025magic,
  title = {Magic Resources of the Heisenberg Picture},
  author = {Dowling, Neil and Kos, Pavel and Turkeshi, Xhek},
  journal = {Phys. Rev. Lett.},
  volume = {135},
  issue = {5},
  pages = {050401},
  numpages = {10},
  year = {2025},
  month = {Jul},
  publisher = {American Physical Society},
  doi = {10.1103/p7xt-s9nz},
  url = {https://link.aps.org/doi/10.1103/p7xt-s9nz}
}

@article{masotllima2024stabilizer,
  title = {Stabilizer Tensor Networks: Universal Quantum Simulator on a Basis of Stabilizer States},
  author = {Masot-Llima, Sergi and Garcia-Saez, Artur},
  journal = {Phys. Rev. Lett.},
  volume = {133},
  issue = {23},
  pages = {230601},
  numpages = {6},
  year = {2024},
  month = {Dec},
  publisher = {American Physical Society},
  doi = {10.1103/PhysRevLett.133.230601},
  url = {https://link.aps.org/doi/10.1103/PhysRevLett.133.230601}
}

@article{aditya2025mpembaeffectsquantumcomplexity,
      title={Mpemba Effects in Quantum Complexity}, 
      author={Sreemayee Aditya and Alessandro Summer and Piotr Sierant and Xhek Turkeshi},
      year={2025},
      journal={arXiv:2509.22176},
      archivePrefix={arXiv},
      primaryClass={quant-ph},
      url={https://arxiv.org/abs/2509.22176}, 
}

@article{hernándezyanes2025nonstabilizernessquantumenhancedmetrologicalprotocols,
      title={Non-stabilizerness in quantum-enhanced metrological protocols}, 
      author={Tanausú Hernández-Yanes and Piotr Sierant and Jakub Zakrzewski and Marcin Płodzień},
      year={2025},
      journal={arXiv:2510.01380},
      archivePrefix={arXiv},
      primaryClass={quant-ph},
      url={https://arxiv.org/abs/2510.01380}, 
}

@article{falcão2025magicdynamicsmanybodylocalized,
      title={Magic dynamics in many-body localized systems}, 
      author={Pedro R. Nicácio Falcão and Piotr Sierant and Jakub Zakrzewski and Emanuele Tirrito},
      year={2025},
      journal={arXiv:2503.07468},
      archivePrefix={arXiv},
      primaryClass={quant-ph},
      url={https://arxiv.org/abs/2503.07468}, 
}

@article{PRXQuantum.4.010301,
  title = {Scalable Measures of Magic Resource for Quantum Computers},
  author = {Haug, Tobias and Kim, M.S.},
  journal = {PRX Quantum},
  volume = {4},
  issue = {1},
  pages = {010301},
  numpages = {23},
  year = {2023},
  month = {Jan},
  publisher = {American Physical Society},
  doi = {10.1103/PRXQuantum.4.010301},
  url = {https://link.aps.org/doi/10.1103/PRXQuantum.4.010301}
}

@article{Oliviero_2022,
  author = {Oliviero, Salvatore F. E. and Leone, Lorenzo and Hamma, Alioscia and Lloyd, Seth},
  title = {Measuring magic on a quantum processor},
  journal = {npj Quantum Information},
  year = {2022},
  date = {2022-12-19},
  volume = {8},
  number = {1},
  pages = {148},
  doi = {10.1038/s41534-022-00666-5},
  url = {https://doi.org/10.1038/s41534-022-00666-5},
  issn = {2056-6387}
}

@article{xorfwht,
      title={A fast and exact approach for stabilizer R\'enyi entropy via the XOR-FWHT algorithm}, 
      author={Xuyang Huang and Han-Ze Li and Jian-Xin Zhong},
      year={2026},
      journal={arXiv:2512.24685},
      archivePrefix={arXiv},
      primaryClass={quant-ph},
      url={https://arxiv.org/abs/2512.24685}, 
}

@article{qiskit,
      title={Quantum computing with Qiskit}, 
      author={Ali Javadi-Abhari and Matthew Treinish and Kevin Krsulich and Christopher J. Wood and Jake Lishman and Julien Gacon and Simon Martiel and Paul D. Nation and Lev S. Bishop and Andrew W. Cross and Blake R. Johnson and Jay M. Gambetta},
      year={2024},
      journal={arXiv:2405.08810},
      archivePrefix={arXiv},
      primaryClass={quant-ph},
      url={https://arxiv.org/abs/2405.08810}, 
}

@article{Zhang_2023,
   title={TensorCircuit: a Quantum Software Framework for the NISQ Era},
   volume={7},
   ISSN={2521-327X},
   url={http://dx.doi.org/10.22331/q-2023-02-02-912},
   DOI={10.22331/q-2023-02-02-912},
   journal={Quantum},
   publisher={Verein zur Forderung des Open Access Publizierens in den Quantenwissenschaften},
   author={Zhang, Shi-Xin and Allcock, Jonathan and Wan, Zhou-Quan and Liu, Shuo and Sun, Jiace and Yu, Hao and Yang, Xing-Han and Qiu, Jiezhong and Ye, Zhaofeng and Chen, Yu-Qin and Lee, Chee-Kong and Zheng, Yi-Cong and Jian, Shao-Kai and Yao, Hong and Hsieh, Chang-Yu and Zhang, Shengyu},
   year={2023},
   month=feb, pages={912} }

@article{tirrito2025universalspreadingnonstabilizernessquantum,
      title={Universal Spreading of Nonstabilizerness and Quantum Transport}, 
      author={Emanuele Tirrito and Poetri Sonya Tarabunga and Devendra Singh Bhakuni and Marcello Dalmonte and Piotr Sierant and Xhek Turkeshi},
      year={2025},
      journal={arXiv:2506.12133},
      archivePrefix={arXiv},
      primaryClass={quant-ph},
      url={https://arxiv.org/abs/2506.12133}, 
}

@article{exp1,
  title = {Quantum Computation with Ions in Thermal Motion},
  author = {S\o{}rensen, Anders and M\o{}lmer, Klaus},
  journal = {Phys. Rev. Lett.},
  volume = {82},
  issue = {9},
  pages = {1971--1974},
  numpages = {0},
  year = {1999},
  month = {Mar},
  publisher = {American Physical Society},
  doi = {10.1103/PhysRevLett.82.1971},
  url = {https://link.aps.org/doi/10.1103/PhysRevLett.82.1971}
}

@article{exp2,
   title={Probing R\'enyi entanglement entropy via randomized measurements},
   volume={364},
   ISSN={1095-9203},
   url={http://dx.doi.org/10.1126/science.aau4963},
   DOI={10.1126/science.aau4963},
   number={6437},
   journal={Science},
   publisher={American Association for the Advancement of Science (AAAS)},
   author={Brydges, Tiff and Elben, Andreas and Jurcevic, Petar and Vermersch, Benoît and Maier, Christine and Lanyon, Ben P. and Zoller, Peter and Blatt, Rainer and Roos, Christian F.},
   year={2019},
   month=apr, pages={260–263} }

@article{QME1,
   title={The quantum Mpemba effects},
   volume={7},
   ISSN={2522-5820},
   url={http://dx.doi.org/10.1038/s42254-025-00838-0},
   DOI={10.1038/s42254-025-00838-0},
   number={8},
   journal={Nature Reviews Physics},
   publisher={Springer Science and Business Media LLC},
   author={Ares, Filiberto and Calabrese, Pasquale and Murciano, Sara},
   year={2025},
   month=jul, pages={451–460} }

@article{QME3,
  title = {Quantum Mpemba Effect in Random Circuits},
  author = {Turkeshi, Xhek and Calabrese, Pasquale and De Luca, Andrea},
  journal = {Phys. Rev. Lett.},
  volume = {135},
  issue = {4},
  pages = {040403},
  numpages = {10},
  year = {2025},
  month = {Jul},
  publisher = {American Physical Society},
  doi = {10.1103/5d6p-8d1b},
  url = {https://link.aps.org/doi/10.1103/5d6p-8d1b}
}

@article{sticlet2025nonstabilizernessopenxxzspin,
      title={Non-stabilizerness in open {XXZ} spin chains: Universal scaling and dynamics}, 
      author={Doru Sticlet and Balázs Dóra and Dominik Szombathy and Gergely Zaránd and Cătălin Paşcu Moca},
      year={2025},
      journal={arXiv:2504.11139},
      archivePrefix={arXiv},
      primaryClass={quant-ph},
      url={https://arxiv.org/abs/2504.11139}, 
}

@article{bera2025nonstabilizernesssachdevyekitaevmodel,
      title={Non-Stabilizerness of Sachdev-Ye-Kitaev Model}, 
      author={Surajit Bera and Marco Schirò},
      year={2025},
      journal={arXiv:2502.01582},
      archivePrefix={arXiv},
      primaryClass={quant-ph},
      url={https://arxiv.org/abs/2502.01582}, 
}

@article{zhang2024quantummagicdynamicsrandom,
      title={Quantum magic dynamics in random circuits}, 
      author={Yuzhen Zhang and Yingfei Gu},
      year={2024},
      journal={arXiv:2410.21128},
      archivePrefix={arXiv},
      primaryClass={quant-ph},
      url={https://arxiv.org/abs/2410.21128}, 
}

@article{szombathy2025independentstabilizerrenyientropy,
      title={Independent stabilizer R\'enyi entropy and entanglement fluctuations in random unitary circuits}, 
      author={Dominik Szombathy and Angelo Valli and Cătălin Paşcu Moca and Lóránt Farkas and Gergely Zaránd},
      year={2025},
      journal={arXiv:2501.11489},
      archivePrefix={arXiv},
      primaryClass={quant-ph},
      url={https://arxiv.org/abs/2501.11489}, 
}

@article{hou2025stabilizerentanglementenhancesmagic,
      title={Stabilizer Entanglement Enhances Magic Injection}, 
      author={Zong-Yue Hou and ChunJun Cao and Zhi-Cheng Yang},
      year={2025},
      journal={arXiv:2503.20873},
      archivePrefix={arXiv},
      primaryClass={quant-ph},
      url={https://arxiv.org/abs/2503.20873}, 
}

@article{hoshino2025stabilizerrenyientropyconformal,
      title={Stabilizer R\'enyi Entropy and Conformal Field Theory}, 
      author={Masahiro Hoshino and Masaki Oshikawa and Yuto Ashida},
      year={2025},
      journal={arXiv:2503.13599},
      archivePrefix={arXiv},
      primaryClass={quant-ph},
      url={https://arxiv.org/abs/2503.13599}, 
}

@article{Haug2023stabilizerentropies,
  doi = {10.22331/q-2023-08-28-1092},
  url = {https://doi.org/10.22331/q-2023-08-28-1092},
  title = {Stabilizer entropies and nonstabilizerness monotones},
  author = {Haug, Tobias and Piroli, Lorenzo},
  journal = {{Quantum}},
  issn = {2521-327X},
  publisher = {{Verein zur F{\"{o}}rderung des Open Access Publizierens in den Quantenwissenschaften}},
  volume = {7},
  pages = {1092},
  month = aug,
  year = {2023}
}

@article{liu2022manybody,
  title = {Many-Body Quantum Magic},
  author = {Liu, Zi-Wen and Winter, Andreas},
  journal = {PRX Quantum},
  volume = {3},
  issue = {2},
  pages = {020333},
  numpages = {18},
  year = {2022},
  month = {May},
  publisher = {American Physical Society},
  doi = {10.1103/PRXQuantum.3.020333},
  url = {https://link.aps.org/doi/10.1103/PRXQuantum.3.020333}
}

@article{Odavic2025PRBStabilizerEntropy,
  title   = {Stabilizer entropy in nonintegrable quantum evolutions},
  author  = {Odavi{\'c}, J. and Viscardi, M. and Hamma, A.},
  journal = {Phys. Rev. B},
  volume  = {112},
  pages   = {104301},
  year    = {2025},
  doi     = {10.1103/y9r6-dx7p}
}

@article{chitambar2019quantum,
  title = {Quantum resource theories},
  author = {Chitambar, Eric and Gour, Gilad},
  journal = {Rev. Mod. Phys.},
  volume = {91},
  issue = {2},
  pages = {025001},
  numpages = {48},
  year = {2019},
  month = {Apr},
  publisher = {American Physical Society},
  doi = {10.1103/RevModPhys.91.025001},
  url = {https://link.aps.org/doi/10.1103/RevModPhys.91.025001}
}

@article{Turkeshi_2024_2,
   title={Magic spreading in random quantum circuits},
   volume={16},
   ISSN={2041-1723},
   url={http://dx.doi.org/10.1038/s41467-025-57704-x},
   DOI={10.1038/s41467-025-57704-x},
   number={1},
pages={2575},
   journal={Nat. Commun.},
   publisher={Springer Science and Business Media LLC},
   author={Turkeshi, Xhek and Tirrito, Emanuele and Sierant, Piotr},
   year={2025},
   month=mar }

@article{Leone_2022,
   title={{Stabilizer Rényi Entropy}},
   volume={128},
   ISSN={1079-7114},
   url={http://dx.doi.org/10.1103/PhysRevLett.128.050402},
   DOI={10.1103/physrevlett.128.050402},
   number={5},
   journal={Physical Review Letters},
pages={050402},
   publisher={American Physical Society (APS)},
   author={Leone, Lorenzo and Oliviero, Salvatore F.E. and Hamma, Alioscia},
   year={2022},
   month=feb
}

@article{gottesman1998heisenbergrepresentationquantumcomputers,
title={The Heisenberg Representation of Quantum Computers}, 
author={Daniel Gottesman},
year={1998},
journal={arXiv:quant-ph/9807006},
archivePrefix={arXiv},
primaryClass={quant-ph},
url={https://arxiv.org/abs/quant-ph/9807006}, 
}

@article{PhysRevA.70.052328,
  title = {Improved simulation of stabilizer circuits},
  author = {Aaronson, Scott and Gottesman, Daniel},
  journal = {Phys. Rev. A},
  volume = {70},
  issue = {5},
  pages = {052328},
  numpages = {14},
  year = {2004},
  month = {Nov},
  publisher = {American Physical Society},
  doi = {10.1103/PhysRevA.70.052328},
  url = {https://link.aps.org/doi/10.1103/PhysRevA.70.052328}
}

@article{LarkinOvchinnikov1969JETP,
  author  = {Larkin, A. I. and Ovchinnikov, Yu. N.},
  title   = {Quasiclassical Method in the Theory of Superconductivity},
  journal = {Soviet Physics JETP},
  volume  = {28},
  number  = {6},
  pages   = {1200--1205},
  year    = {1969},
  month   = jun,
  url     = {https://jetp.ras.ru/cgi-bin/dn/e_028_06_1200.pdf}
}

@article{ShenkerStanford2014JHEP,
  author  = {Shenker, Stephen H. and Stanford, Douglas},
  title   = {Black holes and the butterfly effect},
  journal = {Journal of High Energy Physics},
  volume  = {2014},
  number  = {3},
  pages   = {067},
  year    = {2014},
  doi     = {10.1007/JHEP03(2014)067},
  url     = {https://doi.org/10.1007/JHEP03(2014)067},
}

@article{MaldacenaShenkerStanford2016JHEP,
  author  = {Maldacena, Juan and Shenker, Stephen H. and Stanford, Douglas},
  title   = {A bound on chaos},
  journal = {Journal of High Energy Physics},
  volume  = {2016},
  number  = {8},
  pages   = {106},
  year    = {2016},
  doi     = {10.1007/JHEP08(2016)106},
  url     = {https://doi.org/10.1007/JHEP08(2016)106},
}

@article{Swingle2018NatPhys,
  author  = {Swingle, Brian},
  title   = {Unscrambling the physics of out-of-time-order correlators},
  journal = {Nature Physics},
  volume  = {14},
  number  = {10},
  pages   = {988--990},
  year    = {2018},
  doi     = {10.1038/s41567-018-0295-5},
  url     = {https://doi.org/10.1038/s41567-018-0295-5}
}

@article{HashimotoMurataYoshii2017JHEP,
  author  = {Hashimoto, Koji and Murata, Keiju and Yoshii, Ryosuke},
  title   = {Out-of-time-order correlators in quantum mechanics},
  journal = {Journal of High Energy Physics},
  volume  = {2017},
  number  = {10},
  pages   = {138},
  year    = {2017},
  doi     = {10.1007/JHEP10(2017)138},
  url     = {https://doi.org/10.1007/JHEP10(2017)138},
}

@article{RobertsYoshida2017JHEP,
  author  = {Roberts, Daniel A. and Yoshida, Beni},
  title   = {Chaos and complexity by design},
  journal = {Journal of High Energy Physics},
  volume  = {2017},
  number  = {4},
  pages   = {121},
  year    = {2017},
  doi     = {10.1007/JHEP04(2017)121},
  url     = {https://doi.org/10.1007/JHEP04(2017)121},
}

@article{CotlerHunterJonesLiuYoshida2017JHEP,
  author  = {Cotler, Jordan and Hunter-Jones, Nicholas and Liu, Junyu and Yoshida, Beni},
  title   = {Chaos, complexity, and random matrices},
  journal = {Journal of High Energy Physics},
  volume  = {2017},
  number  = {11},
  pages   = {048},
  year    = {2017},
  doi     = {10.1007/JHEP11(2017)048},
  url     = {https://doi.org/10.1007/JHEP11(2017)048},
}

@article{Entanglement1,
  title = {Entanglement in many-body systems},
  author = {Amico, Luigi and Fazio, Rosario and Osterloh, Andreas and Vedral, Vlatko},
  journal = {Rev. Mod. Phys.},
  volume = {80},
  issue = {2},
  pages = {517--576},
  numpages = {0},
  year = {2008},
  month = {May},
  publisher = {American Physical Society},
  doi = {10.1103/RevModPhys.80.517},
  url = {https://link.aps.org/doi/10.1103/RevModPhys.80.517}
}

@article{Entanglement2,
  title = {Quantum entanglement},
  author = {Horodecki, Ryszard and Horodecki, Pawe\l{} and Horodecki, Micha\l{} and Horodecki, Karol},
  journal = {Rev. Mod. Phys.},
  volume = {81},
  issue = {2},
  pages = {865--942},
  numpages = {0},
  year = {2009},
  month = {Jun},
  publisher = {American Physical Society},
  doi = {10.1103/RevModPhys.81.865},
  url = {https://link.aps.org/doi/10.1103/RevModPhys.81.865}
}

@article{LeoneOlivieroZhouHamma2021Quantum,
  author  = {Leone, Lorenzo and Oliviero, Salvatore F. E. and Zhou, You and Hamma, Alioscia},
  title   = {Quantum Chaos is Quantum},
  journal = {Quantum},
  volume  = {5},
  pages   = {453},
  year    = {2021},
  doi     = {10.22331/q-2021-05-04-453},
  url     = {https://doi.org/10.22331/q-2021-05-04-453},
}

@article{GotoNosakaNozaki2022PRD,
  author  = {Goto, Kanato and Nosaka, Tomonori and Nozaki, Masahiro},
  title   = {Probing chaos by magic monotones},
  journal = {Physical Review D},
  volume  = {106},
  number  = {12},
  pages   = {126009},
  year    = {2022},
  doi     = {10.1103/PhysRevD.106.126009},
  url     = {https://doi.org/10.1103/PhysRevD.106.126009}
}

@article{PaviglianitiLamiColluraSilva2025PRXQ,
  author  = {Paviglianiti, Alessio and Lami, Guglielmo and Collura, Mario and Silva, Alessandro},
  title   = {Estimating Nonstabilizerness Dynamics Without Simulating It},
  journal = {PRX Quantum},
  volume  = {6},
  pages   = {030320},
  year    = {2025},
  month   = aug,
  doi     = {10.1103/msm2-vmg7},
  url     = {https://doi.org/10.1103/msm2-vmg7},
}

@article{HaydenPreskill2007JHEP,
  author  = {Hayden, Patrick and Preskill, John},
  title   = {Black holes as mirrors: quantum information in random subsystems},
  journal = {Journal of High Energy Physics},
  volume  = {2007},
  number  = {9},
  pages   = {120},
  year    = {2007},
  doi     = {10.1088/1126-6708/2007/09/120},
  url     = {https://doi.org/10.1088/1126-6708/2007/09/120},
}

@article{SekinoSusskind2008JHEP,
  author  = {Sekino, Yasuhiro and Susskind, Leonard},
  title   = {Fast Scramblers},
  journal = {Journal of High Energy Physics},
  volume  = {2008},
  number  = {10},
  pages   = {065},
  year    = {2008},
  doi     = {10.1088/1126-6708/2008/10/065},
  url     = {https://doi.org/10.1088/1126-6708/2008/10/065},
}

@article{HosurQiRobertsYoshida2016JHEP,
  author  = {Hosur, Pavan and Qi, Xiao-Liang and Roberts, Daniel A. and Yoshida, Beni},
  title   = {Chaos in quantum channels},
  journal = {Journal of High Energy Physics},
  volume  = {2016},
  number  = {2},
  pages   = {004},
  year    = {2016},
  doi     = {10.1007/JHEP02(2016)004},
  url     = {https://doi.org/10.1007/JHEP02(2016)004},
}

@article{vonKeyserlingkRakovszkyPollmannSondhi2018PRX,
  author  = {von Keyserlingk, Curt W. and Rakovszky, Tibor and Pollmann, Frank and Sondhi, Shivaji L.},
  title   = {Operator Hydrodynamics, {OTOCs}, and Entanglement Growth in Systems without Conservation Laws},
  journal = {Physical Review X},
  volume  = {8},
  pages   = {021013},
  year    = {2018},
  doi     = {10.1103/PhysRevX.8.021013},
  url     = {https://doi.org/10.1103/PhysRevX.8.021013},
}

@article{KhemaniVishwanathHuse2018PRX,
  author  = {Khemani, Vedika and Vishwanath, Ashvin and Huse, David A.},
  title   = {Operator Spreading and the Emergence of Dissipative Hydrodynamics under Unitary Evolution with Conservation Laws},
  journal = {Physical Review X},
  volume  = {8},
  pages   = {031057},
  year    = {2018},
  doi     = {10.1103/PhysRevX.8.031057},
  url     = {https://doi.org/10.1103/PhysRevX.8.031057},
}

@article{LiuZhangYinZhang2024PRL,
  author  = {Liu, Shuo and Zhang, Hao-Kai and Yin, Shuai and Zhang, Shi-Xin},
  title   = {Symmetry Restoration and Quantum Mpemba Effect in Symmetric Random Circuits},
  journal = {Physical Review Letters},
  volume  = {133},
  pages   = {140405},
  year    = {2024},
  doi     = {10.1103/PhysRevLett.133.140405},
  url     = {https://doi.org/10.1103/PhysRevLett.133.140405},
}

@article{LiuZhangYinZhangYao2024arXiv,
   title={Symmetry restoration and quantum Mpemba effect in many-body localization systems},
   volume={70},
   ISSN={2095-9273},
   url={http://dx.doi.org/10.1016/j.scib.2025.10.017},
   DOI={10.1016/j.scib.2025.10.017},
   number={23},
   journal={Science Bulletin},
   publisher={Elsevier BV},
   author={Liu, Shuo and Zhang, Hao-Kai and Yin, Shuai and Zhang, Shi-Xin and Yao, Hong},
   year={2025},
   month=dec, pages={3991–3996} }

@article{YuLiZhang2025CPL,
  author  = {Yu, Hui and Li, Zi-Xiang and Zhang, Shi-Xin},
  title   = {Symmetry Breaking Dynamics in Quantum Many-Body Systems},
  journal = {Chinese Physics Letters},
  volume  = {42},
  number  = {11},
  pages   = {110602},
  year    = {2025},
  doi     = {10.1088/0256-307X/42/11/110602},
  url     = {https://doi.org/10.1088/0256-307X/42/11/110602}
}

@article{DankertCleveEmersonLivine2006arXiv,
  title = {Exact and approximate unitary 2-designs and their application to fidelity estimation},
  author = {Dankert, Christoph and Cleve, Richard and Emerson, Joseph and Livine, Etera},
  journal = {Phys. Rev. A},
  volume = {80},
  issue = {1},
  pages = {012304},
  numpages = {6},
  year = {2009},
  month = {Jul},
  publisher = {American Physical Society},
  doi = {10.1103/PhysRevA.80.012304},
  url = {https://link.aps.org/doi/10.1103/PhysRevA.80.012304}
}

@article{DankertCleveEmersonLivine2009PRA,
  author  = {Dankert, Christoph and Cleve, Richard and Emerson, Joseph and Livine, Etera},
  title   = {Exact and approximate unitary 2-designs and their application to fidelity estimation},
  journal = {Physical Review A},
  volume  = {80},
  number  = {1},
  pages   = {012304},
  year    = {2009},
  doi     = {10.1103/PhysRevA.80.012304},
  url     = {https://doi.org/10.1103/PhysRevA.80.012304}
}

@article{HarrowLow2009CMP,
  author  = {Harrow, Aram W. and Low, Richard A.},
  title   = {Random Quantum Circuits are Approximate 2-designs},
  journal = {Communications in Mathematical Physics},
  volume  = {291},
  pages   = {257--302},
  year    = {2009},
  doi     = {10.1007/s00220-009-0873-6},
  url     = {https://doi.org/10.1007/s00220-009-0873-6},
}

@article{BrandaoHarrowHorodecki2016CMP,
  author  = {Brand{\~a}o, Fernando G. S. L. and Harrow, Aram W. and Horodecki, Micha{\l}},
  title   = {Local Random Quantum Circuits are Approximate Polynomial-Designs},
  journal = {Communications in Mathematical Physics},
  volume  = {346},
  number  = {2},
  pages   = {397--434},
  year    = {2016},
  doi     = {10.1007/s00220-016-2706-8},
  url     = {https://doi.org/10.1007/s00220-016-2706-8},
}

@article{HoChoi2022PRL,
  author  = {Ho, Wen Wei and Choi, Soonwon},
  title   = {Exact Emergent Quantum State Designs from Quantum Chaotic Dynamics},
  journal = {Physical Review Letters},
  volume  = {128},
  pages   = {060601},
  year    = {2022},
  doi     = {10.1103/PhysRevLett.128.060601},
  url     = {https://doi.org/10.1103/PhysRevLett.128.060601},
}

@article{IppolitiHo2023PRXQ,
  author  = {Ippoliti, Matteo and Ho, Wen Wei},
  title   = {Dynamical Purification and the Emergence of Quantum State Designs from the Projected Ensemble},
  journal = {PRX Quantum},
  volume  = {4},
  pages   = {030322},
  year    = {2023},
  doi     = {10.1103/PRXQuantum.4.030322},
  url     = {https://doi.org/10.1103/PRXQuantum.4.030322},
}

@article{ClaeysLamacraftVicary2024JPhysA,
  author  = {Claeys, Pieter W. and Lamacraft, Austen and Vicary, Jamie},
  title   = {From dual-unitary to biunitary: a 2-categorical model for exactly-solvable many-body quantum dynamics},
  journal = {Journal of Physics A: Mathematical and Theoretical},
  volume  = {57},
  number  = {33},
  pages   = {335301},
  year    = {2024},
  doi     = {10.1088/1751-8121/ad653f},
  url     = {https://doi.org/10.1088/1751-8121/ad653f},
}

@article{Prosen2021Chaos,
  author  = {Prosen, Toma{\v{z}}},
  title   = {Many-body quantum chaos and dual-unitarity round-a-face},
  journal = {Chaos},
  volume  = {31},
  number  = {9},
  pages   = {093101},
  year    = {2021},
  doi     = {10.1063/5.0056970},
  url     = {https://doi.org/10.1063/5.0056970},
}

@article{HuangLiHuseChan2023Scholarpedia,
  author  = {Huang, Ke and Li, Xiao and Huse, David A. and Chan, Amos},
  title   = {Out-of-time-order correlations and quantum chaos},
  journal = {Scholarpedia},
  volume  = {18},
  number  = {4},
  pages   = {55237},
  year    = {2023},
  doi     = {10.4249/scholarpedia.55237},
  url     = {http://www.scholarpedia.org/article/Out-of-time-order_correlations_and_quantum_chaos}
}

@article{bnld-2chd,
  title = {Emergent Unitary Designs for Encoded Qubits from Coherent Errors and Syndrome Measurements},
  author = {Cheng, Zihan and Huang, Eric and Khemani, Vedika and Gullans, Michael J. and Ippoliti, Matteo},
  journal = {PRX Quantum},
  volume = {6},
  issue = {3},
  pages = {030333},
  numpages = {22},
  year = {2025},
  month = {Aug},
  publisher = {American Physical Society},
  doi = {10.1103/bnld-2chd},
  url = {https://link.aps.org/doi/10.1103/bnld-2chd}
}

@article{PhysRevResearch.7.L012011,
  title = {Solvable entanglement dynamics in quantum circuits with generalized space-time duality},
  author = {Liu, Chuan and Ho, Wen Wei},
  journal = {Phys. Rev. Res.},
  volume = {7},
  issue = {1},
  pages = {L012011},
  numpages = {6},
  year = {2025},
  month = {Jan},
  publisher = {American Physical Society},
  doi = {10.1103/PhysRevResearch.7.L012011},
  url = {https://link.aps.org/doi/10.1103/PhysRevResearch.7.L012011}
}

@article{Abanin2019RMP,
  author  = {Abanin, Dmitry A. and Altman, Ehud and Bloch, Immanuel and Serbyn, Maksym},
  title   = {Colloquium: Many-body localization, thermalization, and entanglement},
  journal = {Reviews of Modern Physics},
  volume  = {91},
  pages   = {021001},
  year    = {2019},
  doi     = {10.1103/RevModPhys.91.021001}
}

@article{NandkishoreHuse2015,
  author  = {Nandkishore, Rahul and Huse, David A.},
  title   = {Many-Body Localization and Thermalization in Quantum Statistical Mechanics},
  journal = {Annual Review of Condensed Matter Physics},
  volume  = {6},
  pages   = {15--38},
  year    = {2015},
  doi     = {10.1146/annurev-conmatphys-031214-014726}
}

@article{Serbyn2013LIOM,
  title   = {Local Conservation Laws and the Structure of the Many-Body Localized States},
  author  = {Serbyn, Maksym and Papi{\'c}, Z. and Abanin, Dmitry A.},
  journal = {Phys. Rev. Lett.},
  volume  = {111},
  pages   = {127201},
  year    = {2013},
  doi     = {10.1103/PhysRevLett.111.127201}
}

@article{AletLaflorencie2018,
  author  = {Alet, Fabien and Laflorencie, Nicolas},
  title   = {Many-body localization: An introduction and selected topics},
  journal = {Comptes Rendus Physique},
  volume  = {19},
  number  = {6},
  pages   = {498--525},
  year    = {2018},
  doi     = {10.1016/j.crhy.2018.03.003}
}

@article{Schulz2019PRL,
  author  = {Schulz, M. and Hooley, C. A. and Moessner, R. and Pollmann, F.},
  title   = {Stark Many-Body Localization},
  journal = {Physical Review Letters},
  volume  = {122},
  pages   = {040606},
  year    = {2019},
  doi     = {10.1103/PhysRevLett.122.040606}
}

@article{Sala2020PRXFragmentation,
  title   = {Ergodicity Breaking Arising from Hilbert Space Fragmentation in Dipole-Conserving Hamiltonians},
  author  = {Sala, Pablo and Rakovszky, Tibor and Verresen, Ruben and Knap, Michael and Pollmann, Frank},
  journal = {Phys. Rev. X},
  volume  = {10},
  pages   = {011047},
  year    = {2020},
  doi     = {10.1103/PhysRevX.10.011047}
}

@article{vanNieuwenburg2019PNAS,
author = {Evert van Nieuwenburg  and Yuval Baum  and Gil Refael },
title = {From Bloch oscillations to many-body localization in clean interacting systems},
journal = {Proceedings of the National Academy of Sciences},
volume = {116},
number = {19},
pages = {9269-9274},
year = {2019},
doi = {10.1073/pnas.1819316116},
URL = {https://www.pnas.org/doi/abs/10.1073/pnas.1819316116},}

@article{Moudgalya_2022,
   title={Quantum many-body scars and Hilbert space fragmentation: a review of exact results},
   volume={85},
   ISSN={1361-6633},
   url={http://dx.doi.org/10.1088/1361-6633/ac73a0},
   DOI={10.1088/1361-6633/ac73a0},
   number={8},
   journal={Reports on Progress in Physics},
   publisher={IOP Publishing},
   author={Moudgalya, Sanjay and Bernevig, B Andrei and Regnault, Nicolas},
   year={2022},
   month=jul, pages={086501} }

@article{Scherg2021NatCommun,
  author  = {Scherg, Sebastian and Kohlert, Thomas and Sala, Pablo and Pollmann, Frank and Bharath, H. M. and Bloch, Immanuel and Aidelsburger, Monika},
  title   = {Observing non-ergodicity due to kinetic constraints in tilted Fermi-Hubbard chains},
  journal = {Nature Communications},
  volume  = {12},
  pages   = {4490},
  year    = {2021},
  doi     = {10.1038/s41467-021-24726-0}
}

@article{Sala2020PRX,
  author  = {Sala, Pablo and Rakovszky, Tibor and Verresen, Ruben and Knap, Michael and Pollmann, Frank},
  title   = {Ergodicity-breaking arising from Hilbert space fragmentation in dipole-conserving Hamiltonians},
  journal = {Physical Review X},
  volume  = {10},
  pages   = {011047},
  year    = {2020},
  doi     = {10.1103/PhysRevX.10.011047}
}

@article{Khemani2020PRB,
  author  = {Khemani, Vedika and Hermele, Michael and Nandkishore, Rahul M.},
  title   = {Localization from Hilbert space shattering: from theory to physical realizations},
  journal = {Physical Review B},
  volume  = {101},
  pages   = {174204},
  year    = {2020},
  doi     = {10.1103/PhysRevB.101.174204}
}

@article{Nandy2024PRB,
  author  = {Nandy, S. and Herbrych, J. and Lenar{\v{c}}i{\v{c}}, Z. and G{\l}{\'o}dkowski, A. and Prelov{\v{s}}ek, P. and Mierzejewski, M.},
  title   = {Emergent dipole moment conservation and subdiffusion in tilted chains},
  journal = {Physical Review B},
  volume  = {109},
  pages   = {115120},
  year    = {2024},
  doi     = {10.1103/PhysRevB.109.115120}
}

@article{Gluck2002PhysRep,
  author  = {Gl{\"u}ck, M. and Kolovsky, A. R. and Korsch, H. J.},
  title   = {Wannier--Stark resonances in optical and semiconductor superlattices},
  journal = {Physics Reports},
  volume  = {366},
  number  = {3},
  pages   = {103--182},
  year    = {2002},
  doi     = {10.1016/S0370-1573(02)00142-4}
}

@article{Adler2024Nature,
  author  = {Adler, Daniel and Wei, David and Will, Melissa and Srakaew, Kritsana and Agrawal, Suchita and Weckesser, Pascal and Moessner, Roderich and Pollmann, Frank and Bloch, Immanuel and Zeiher, Johannes},
  title   = {Observation of Hilbert space fragmentation and fractonic excitations in 2D},
  journal = {Nature},
  volume  = {636},
  pages   = {80--85},
  year    = {2024},
  doi     = {10.1038/s41586-024-08188-0}
}

@article{Brenes2018PRL,
  author  = {Brenes, Marlon and Dalmonte, Marcello and Heyl, Markus and Scardicchio, Antonello},
  title   = {Many-body localization dynamics from gauge invariance},
  journal = {Physical Review Letters},
  volume  = {120},
  pages   = {030601},
  year    = {2018},
  doi     = {10.1103/PhysRevLett.120.030601}
}

@article{Karpov2021PRL,
  author  = {Karpov, P. and Verdel, R. and Huang, Y.-P. and Schmitt, M. and Heyl, M.},
  title   = {Disorder-free localization in an interacting two-dimensional lattice gauge theory},
  journal = {Physical Review Letters},
  volume  = {126},
  pages   = {130401},
  year    = {2021},
  doi     = {10.1103/PhysRevLett.126.130401}
}

@article{sxz1,
  title = {Universal Properties of Many-Body Localization Transitions in Quasiperiodic Systems},
  author = {Zhang, Shi-Xin and Yao, Hong},
  journal = {Phys. Rev. Lett.},
  volume = {121},
  issue = {20},
  pages = {206601},
  numpages = {5},
  year = {2018},
  month = {Nov},
  publisher = {American Physical Society},
  doi = {10.1103/PhysRevLett.121.206601},
  url = {https://link.aps.org/doi/10.1103/PhysRevLett.121.206601}
}

@article{sxz2,
  title = {Discrete Time Crystal Enabled by Stark Many-Body Localization},
  author = {Liu, Shuo and Zhang, Shi-Xin and Hsieh, Chang-Yu and Zhang, Shengyu and Yao, Hong},
  journal = {Phys. Rev. Lett.},
  volume = {130},
  issue = {12},
  pages = {120403},
  numpages = {8},
  year = {2023},
  month = {Mar},
  publisher = {American Physical Society},
  doi = {10.1103/PhysRevLett.130.120403},
  url = {https://link.aps.org/doi/10.1103/PhysRevLett.130.120403}
}

@article{sxz3,
  title = {Probing many-body localization by excited-state variational quantum eigensolver},
  author = {Liu, Shuo and Zhang, Shi-Xin and Hsieh, Chang-Yu and Zhang, Shengyu and Yao, Hong},
  journal = {Phys. Rev. B},
  volume = {107},
  issue = {2},
  pages = {024204},
  numpages = {8},
  year = {2023},
  month = {Jan},
  publisher = {American Physical Society},
  doi = {10.1103/PhysRevB.107.024204},
  url = {https://link.aps.org/doi/10.1103/PhysRevB.107.024204}
}

@article{sxz4,
title = {Symmetry restoration and quantum Mpemba effect in many-body localization systems},
journal = {Science Bulletin},
volume = {70},
number = {23},
pages = {3991-3996},
year = {2025},
issn = {2095-9273},
doi = {https://doi.org/10.1016/j.scib.2025.10.017},
url = {https://www.sciencedirect.com/science/article/pii/S2095927325010394},
author = {Shuo Liu and Hao-Kai Zhang and Shuai Yin and Shi-Xin Zhang and Hong Yao}
}

@article{Yu_2025,
   title={Quantum Mpemba effects from symmetry perspectives},
   volume={35},
   pages={17},
   ISSN={2309-4710},
   url={http://dx.doi.org/10.1007/s43673-025-00157-7},
   number={1},
   journal={AAPPS Bulletin},
   publisher={Springer Science and Business Media LLC},
   author={Yu, Hui and Liu, Shuo and Zhang, Shi-Xin},
   year={2025},
   month=jul }

@article{PhysRev.117.432,
  title = {Wave Functions and Effective Hamiltonian for Bloch Electrons in an Electric Field},
  author = {Wannier, Gregory H.},
  journal = {Phys. Rev.},
  volume = {117},
  issue = {2},
  pages = {432--439},
  numpages = {0},
  year = {1960},
  month = {Jan},
  publisher = {American Physical Society},
  doi = {10.1103/PhysRev.117.432},
  url = {https://link.aps.org/doi/10.1103/PhysRev.117.432}
}

@article{aditya2025growthspreadingquantumresources,
      title={Growth and spreading of quantum resources under random circuit dynamics}, 
      author={Sreemayee Aditya and Xhek Turkeshi and Piotr Sierant},
      year={2025},
      journal={arXiv:2512.14827},
      archivePrefix={arXiv},
      primaryClass={quant-ph},
      url={https://arxiv.org/abs/2512.14827}, 
}

@article{xhek_QME1,
      title={A resource theoretical unification of Mpemba effects: classical and quantum}, 
      author={Alessandro Summer and Mattia Moroder and Laetitia P. Bettmann and Xhek Turkeshi and Iman Marvian and John Goold},
      year={2025},
      journal={arXiv:2507.16976},
      archivePrefix={arXiv},
      primaryClass={quant-ph},
      url={https://arxiv.org/abs/2507.16976}, 
}

@article{xhek_QME2,
      title={Mpemba Effects in Quantum Complexity}, 
      author={Sreemayee Aditya and Alessandro Summer and Piotr Sierant and Xhek Turkeshi},
      year={2025},
      journal={arXiv:2509.22176},
      archivePrefix={arXiv},
      primaryClass={quant-ph},
      url={https://arxiv.org/abs/2509.22176}, 
}

@article{xhek_QME3,
   title={Measurement-Induced Symmetry Restoration and Quantum Mpemba Effect},
   volume={27},
   ISSN={1099-4300},
   url={http://dx.doi.org/10.3390/e27040407},
   DOI={10.3390/e27040407},
   number={4},
   journal={Entropy},
   publisher={MDPI AG},
   author={Di Giulio, Giuseppe and Turkeshi, Xhek and Murciano, Sara},
   year={2025},
   month=apr, pages={407} }

@article{li2025quantummpembaeffectlongranged,
      title={Quantum Mpemba effect in long-ranged U(1)-symmetric random circuits}, 
      author={Han-Ze Li and Ching Hua Lee and Shuo Liu and Shi-Xin Zhang and Jian-Xin Zhong},
      year={2025},
      journal={arXiv:2512.06775},
      archivePrefix={arXiv},
      primaryClass={quant-ph},
      url={https://arxiv.org/abs/2512.06775}, 
}

@article{lhz2,
      title={Non-Hermitian many-body localization in asymmetric chains with long-range interaction}, 
      author={Wen Wang and Han-Ze Li and Jian-Xin Zhong},
      year={2025},
      journal={arXiv:2510.08277},
      archivePrefix={arXiv},
      primaryClass={cond-mat.dis-nn},
      url={https://arxiv.org/abs/2510.08277}, 
}

\clearpage
\newpage
\onecolumngrid

\appendix

% \section{Nonstabilizerness in Wannier-Stark localization}

\section{Analysis for the magic dynamics of SMBL}\label{sec:Analytical-details}
In this appendix, we provide the derivations underlying the strong-tilt analysis in the main text. Starting from Eq.~\eqref{eq:H_stark_tfim}, we write the Hamiltonian as $H=H_0+V$ with
\begin{align}
H_0=\sum_{j=0}^{L-1}\!\Big(F\,j\,Z_j + J\,Z_j Z_{j+1}\Big),\quad
V=h\sum_{j=0}^{L-1} X_j ,
\end{align}
and work in the strong-tilt regime $F\gg J,h$. We construct a Schrieffer--Wolff (SW) effective Hamiltonian $H_{\mathrm{eff}}=e^{S}He^{-S}$ with an anti-Hermitian generator $S^\dagger=-S$. Using the Baker--Campbell--Hausdorff expansion $e^{S}Ae^{-S}=A+[S,A]+\frac{1}{2}[S,[S,A]]+\cdots$ and the series $S=S^{(1)}+O(h^2)$ with $S^{(1)}=O(h)$, truncation to second order yields
\begin{align}
H_{\mathrm{eff}}
= H_0 + V + [S^{(1)},H_0] + \frac{1}{2}[S^{(1)},V] + O(h^3).
\label{eq:Heff_BCH}
\end{align}
The SW condition eliminates the leading off-diagonal contribution in the eigenbasis of $H_0$,
\begin{align}
[H_0,S^{(1)}]=V.
\label{eq:SW_condition}
\end{align}

To solve Eq.~\eqref{eq:SW_condition}, define the diagonal operator $\Delta_j\equiv Fj+J(Z_{j-1}+Z_{j+1})$, with the boundary convention $Z_{-1}=Z_{L}=0$ for open boundary conditions. Using $[Z_j,Y_j]=2iX_j$, one obtains $[H_0,Y_j]=-2i\Delta_j X_j$, and a convenient choice is $S^{(1)}=\sum_j \frac{i h}{2}\Delta_j^{-1}Y_j$. Substituting into Eq.~\eqref{eq:Heff_BCH} gives the standard second-order result
\begin{align}
H_{\mathrm{eff}}
= H_0 + \frac{1}{2}[S^{(1)},V] + O(h^3)
= H_0 + \sum_{j=0}^{L-1}\frac{h^2}{2}\,\Delta_j^{-1} Z_j + O(h^3).
\label{eq:Heff_second_order}
\end{align}

The long-time dephasing dynamics are governed by higher-order diagonal multi-spin terms generated by the SW procedure. To make this structure explicit, we project $H_{\mathrm{eff}}$ onto its diagonal part in the $Z$-product basis $\{\ket{\mathbf{s}}\}$,
$H_{\mathrm{eff}}^{(d)}\equiv\sum_{\mathbf{s}}\ket{\mathbf{s}}\bra{\mathbf{s}}H_{\mathrm{eff}}\ket{\mathbf{s}}\bra{\mathbf{s}}$.
Introduce the diagonal operator basis $Z_S\equiv\prod_{i\in S} Z_i$ for subsets $S\subseteq\{0,\ldots,L-1\}$. Since $\{Z_S\}$ is orthogonal under the Hilbert--Schmidt inner product, $H_{\mathrm{eff}}^{(d)}$ admits the unique expansion
\begin{align}
H_{\mathrm{eff}}^{(d)}=\sum_{S\subseteq\{0,\ldots,L-1\}} c_S\,Z_S,\quad
c_S=2^{-L}\,\Tr\!\big(H_{\mathrm{eff}}^{(d)} Z_S\big).
\label{eq:diag_expansion}
\end{align}
We define the induced diagonal two-body coefficients as $\tilde{J}_{ij}\equiv c_{\{i,j\}}$ and quantify the typical diagonal coupling strength at separation $r$ by the root-mean-square average
\begin{align}
J_{\mathrm{eff}}(r)\equiv
\left(\frac{1}{L-r}\sum_{i=0}^{L-1-r}\tilde{J}_{i,i+r}^{\,2}\right)^{1/2}.
\label{eq:Jeff_def}
\end{align}

In the strong-tilt regime, a diagonal coupling spanning distance $r$ arises from a minimal chain of virtual transverse-field processes. A scaling estimate gives a numerator of order $h^r$, while the intermediate energy denominators are dominated by the Stark ladder and accumulate approximately linearly, $E_i-E_{k_m}\sim mF$ for $m=1,\ldots,r-1$, leading to $\prod_{m=1}^{r-1}(E_i-E_{k_m})\sim F^{\,r-1}(r-1)!$. Absorbing path-counting factors and subleading $O(J/F)$ corrections into a non-universal prefactor $J_0$ (with $J_0\sim h$), we obtain the factorially suppressed scaling form
\begin{align}
J_{\mathrm{eff}}(r)\sim J_0\,\frac{(h/F)^{r-1}}{(r-1)!}.
\label{eq:Jeff_scaling}
\end{align}

We define the dephasing front $r(t)$ by the phase-accumulation condition $t\,J_{\mathrm{eff}}(r(t))\sim 1$. Writing $n\equiv r-1$ and using Stirling's approximation $\ln n!\simeq n\ln n-n$ gives
$n\ln\!\big(n/(e\,h/F)\big)\simeq \ln(tJ_0)$.
Solving in terms of the principal branch $W_0$ of the Lambert $W$ function, defined by $W_0(x)e^{W_0(x)}=x$, yields
\begin{align}
r(t)\simeq 1+\frac{\ln(tJ_0)}{W_0\!\Big(\dfrac{\ln(tJ_0)}{e\,h/F}\Big)}.
\label{eq:r_of_t}
\end{align}

Finally, for the 2-SRE $M_2(t)$ in the dephasing-dominated regime, we use a minimal saturating closure that grows linearly at small argument and approaches a finite-size plateau $M_{\mathrm{sat}}$ at long times,
\begin{align}
M_2(t)\simeq M_{\mathrm{sat}}\tanh\!\big(\gamma\,r(t)\big),
\label{eq:M2_closure_appendix}
\end{align}
where $\gamma$ is a growth coefficient and $J_0\sim h$ sets the microscopic time scale.

\section{Numerical details}\label{app:numerics}
We simulate the unitary dynamics $\ket{\Psi(t)}=e^{-iHt}\ket{\Psi_0}$ of Eq.~\eqref{eq:H_stark_tfim} in the computational ($Z$) basis. Unless stated otherwise, we set $J=1$ and $h=1$, and report time in units of $J^{-1}$. For each $(L,F)$, we obtain $\ket{\Psi(t)}$ by exact diagonalization and spectral time evolution. State normalization is monitored throughout and remains unity, with negligible floating-point error, over the time windows shown in Figs.~\ref{fig1}--\ref{fig3}. For the random product-state ensemble we sample $\ket{\Psi_0}=\bigotimes_{j=0}^{L-1}\ket{\psi_j}$ with each $\ket{\psi_j}$ drawn independently from the Haar measure on the Bloch sphere (implemented by normalizing a complex Gaussian vector in $\mathbb{C}^2$). Results are averaged over $N_{\mathrm{rnd}}$ realizations (we use $N_{\mathrm{rnd}}=100$ unless stated otherwise), with fluctuations estimated from the sample variance or bootstrap resampling.

To evaluate the 2-SRE $M_2(t)$, we compute the Pauli fourth moment in Eq.~\eqref{eq:SRE_def} without explicit enumeration of the $4^L$ Pauli strings. Instead, we use an exact XOR--FWHT method~\cite{xorfwht}: the computation is reformulated in a bitstring language and reduced to $2^L$ fast Walsh--Hadamard transforms acting on autocorrelation vectors generated by XOR shifts of the state amplitudes. This yields a deterministic evaluation with runtime scaling $O(L4^L)$ and natural parallelism over the shift index and over random initial-state realizations.

Long-time plateau values are obtained either by fitting $M_2(t)$ to Eq.~\eqref{eq:log_growth} over a specified late-time window or by a direct late-time average; the chosen windows are stated in the corresponding figure captions. For the finite-size collapse in Fig.~\ref{fig3}, we use $\Delta M_2(F,L)=M_2^{\mathrm{Haar}}(L)-M_2(F,L)$ evaluated from the plateau estimator and extract $(F_c,\nu)$ by least-squares collapse, with uncertainties estimated by bootstrap/jackknife checks.

We validate the implementation by verifying $M_2=0$ for stabilizer product states and by reproducing the Haar benchmark $M_2^{\mathrm{Haar}}(L)$ in Eq.~\eqref{eq:M2_haar} for Haar-random pure states within statistical uncertainty.

\section{Experimental accessibility}\label{app:exp_access}
We propose an experimental protocol on a linear trapped-ion chain to digitally simulate the transverse-field Ising dynamics in the presence of a purely linear Stark tilt, and to measure, as functions of time, both the half-chain 2-R\'enyi entanglement entropy and SRE at $k\!=\!2$ from a standard data set. The platform encodes $L$ qubits in $L$ ions, with each qubit represented by two long-lived internal states $|0\rangle$ and $|1\rangle$. The computational basis $\{|0\rangle,|1\rangle\}$ is fixed by state-dependent fluorescence detection: one of the two levels is coupled to a closed cycling transition and yields bright fluorescence under resonant illumination, whereas the other level remains dark, such that ion-resolved photon counts implement a projective measurement in the $Z$ basis and return a bitstring $s\!\in\!\{0,1\}^L$. The ions are indexed along the chain as $i\!=\!0,\dots,L-1$, and this ordering defines the spatial profile of the linear tilt.

The target Hamiltonian we aim to implement is
\begin{align}
H&=
\sum_{i<j}J_{ij} Z_i Z_j
+
h\sum_{i=0}^{L-1}X_i
+
F\sum_{i=0}^{L-1} i\, Z_i,
\end{align}
where $J_{ij}$ denotes the effective long-range Ising couplings, $h$ is the transverse-field strength, and $F$ is the gradient that generates the linear Stark tilt. We propose to sample the dynamics at discrete times $t_k\!=\!k\,\delta t$, where $\delta t$ is a digital time step and $k$ is a nonnegative integer. A calibration stage is included to establish the mapping from experimental controls to the model parameters. In particular, the transverse-field amplitude can be calibrated via single-qubit Rabi oscillations, which determine the pulse-area-to-angle conversion and thereby fix the implementation of $\theta\!=\!2h\,\delta t$. The coupling matrix $J_{ij}$ can be calibrated by enabling the global entangling interaction for controlled durations and extracting accumulated two-body phases or correlation signals; if appropriate, a compact parametrization may be obtained by fitting to a power-law form. The tilt gradient $F$ can be calibrated by ion-resolved Ramsey phase accumulation, since the extracted $Z$ frequency shifts scale linearly with the ion index $i$; any deviation from strict linearity can be recorded as a systematic uncertainty.

We propose to implement the time evolution using a digital simulation based on a second-order Strang splitting. In a standard trapped-ion implementation, a long-range $XX$ interaction is naturally available via a M\o lmer-S\o rensen mechanism~\cite{exp1},
\begin{align}
U_{XX}(\tau)
&=
\exp\Bigl[-i\tau\sum_{i<j}J_{ij}X_iX_j\Bigr] .
\end{align}
A global basis change maps the native interaction to an effective $ZZ$ interaction,
\begin{align}
U_{ZZ}(\tau)
&=
\Bigl[\prod_{i=0}^{L-1} R_y^{(i)}\!\left(\frac{\pi}{2}\right)\Bigr]\,
U_{XX}(\tau)\,
\Bigl[\prod_{i=0}^{L-1} R_y^{(i)}\!\left(-\frac{\pi}{2}\right)\Bigr] .
\end{align}
The transverse-field step is realized as
\begin{align}
U_X(\delta t)
&=
\prod_{i=0}^{L-1} R_x^{(i)}(\theta),
\end{align}
where $\theta\!=\!2h\,\delta t$.
The linear Stark tilt generates a diagonal evolution over one time step,
\begin{align}
U_F(\delta t)
&=
\exp\Bigl[-i\delta t\,F\sum_{i=0}^{L-1} i\,Z_i\Bigr]
=
\prod_{i=0}^{L-1} R_z^{(i)}(\phi_i),
\end{align}
here $\phi_i=2Fi\,\delta t$ and $\prod_i R_z^{(i)}(\phi_i)$ may be implemented by ion-resolved detunings or AC-Stark shifts that accumulate phases, or equivalently by phase-frame updates that account for the same unitary action in the circuit description via the angles $\phi_i$.

We propose to combine the two-body $ZZ$ term and the linear tilt into a single diagonal block, and define the Strang step as
\begin{align}
U_{\mathrm{Strang}}(\delta t)
&=
U_{\mathrm{diag}}\!\left(\frac{\delta t}{2}\right)\,
U_X(\delta t)\,
U_{\mathrm{diag}}\!\left(\frac{\delta t}{2}\right),
\\
U_{\mathrm{diag}}\!\left(\frac{\delta t}{2}\right)
&=
U_{ZZ}\!\left(\frac{\delta t}{2}\right)\,
\prod_{i=0}^{L-1} R_z^{(i)}\!\left(\frac{\phi_i}{2}\right) .
\end{align}
The evolution to time $t_k$ is then implemented as
\begin{align}
U(t_k)
&\approx
\Bigl[U_{\mathrm{Strang}}(\delta t)\Bigr]^k,
\end{align}
with $t_k\!=\!k\,\delta t$. At the gate level, one Strang step consists of an interaction half-step realizing $U_{ZZ}(\delta t/2)$ through a global $R_y$ basis change surrounding a global entangling window $U_{XX}(\delta t/2)$, together with a linear-phase half-step $\prod_i R_z^{(i)}(\phi_i/2)$, followed by the transverse-field step $\prod_i R_x^{(i)}(\theta)$, and finally the same linear-phase half-step and interaction half-step applied in reverse order to enforce the Strang symmetry.

We propose to prepare four families of product initial states in the main text in order to probe both structured and typical dynamics. The pictorial representation of the architecture is shown in Fig.~\ref{fig4}. The $Z$-polarized state is $|\Psi_Z\rangle$. The $X$-polarized state is $|\Psi_X\rangle\!=\!|+\rangle^{\otimes L}$ with $|+\rangle\!=\!\frac{1}{\sqrt2}(|0\rangle+|1\rangle)$, prepared by a global $R_y(\pi/2)$ rotation on $|0\rangle^{\otimes L}$. The $Y$-polarized state is $|\Psi_Y\rangle\!=\!|+_y\rangle^{\otimes L}$ with $|+_y\rangle\!=\!\frac{1}{\sqrt2}(|0\rangle+i|1\rangle)$, prepared by a global $R_x(-\pi/2)$ rotation. Finally, we propose a fully random Bloch-sphere ensemble implemented as a set $\{|\Psi_{\mathrm{rnd}}^{(r)}\rangle\}_{r=1}^{N_{\mathrm{rnd}}}$ of Haar-uniform random product states, where each ion independently samples $\varphi\!\in\![0,2\pi)$ uniformly and samples $\cos\theta$ uniformly in $[-1,1]$, and is prepared via $R_z(\varphi)R_y(\theta)|0\rangle$ to realize $ |\psi(\theta,\varphi)\rangle\!=\!
\cos\frac{\theta}{2}\,|0\rangle+
e^{i\varphi}\sin\frac{\theta}{2}\,|1\rangle $.
In the random-initial-state case, we propose to repeat the full protocol for each sample index $r$ and report observables averaged over $r$ to suppress finite-size fluctuations and access typical behavior.

To extract both the half-chain 2-R\'enyi entanglement entropy and the $M_2$ dynamics from a common data set, inspired by Ref.~\cite{Oliviero_2022}, we propose a local randomized measurement protocol that requires only single-qubit Clifford operations followed by computational-basis readout. For each time point $t_k$, we sample $N_U$ local single-qubit Clifford unitaries
\begin{align}
C^{(m)}
&=
\bigotimes_{i=0}^{L-1} c_i^{(m)},
\end{align}
with $c_i^{(m)}\!\in\!\mathcal C_1$ and $m\!=\!1,2,\dots,N_U$
apply $C^{(m)}$ to the evolved state $\rho(t_k)$, and then measure all qubits in the computational basis $N_M$ times. This yields bitstrings $s^{(m)}_\alpha\!\in\!\{0,1\}^L$ with $\alpha\!=\!1,2,\dots,N_M$. For the half-chain analysis, we denote by $s^{(m)}_{\alpha,A}$ the restriction of $s^{(m)}_\alpha$ to the subsystem $A\!=\!\{0,\dots,L-1\}$. Conceptually, each $C^{(m)}$ maps computational-basis readout to a random local Pauli measurement basis, and the local single-qubit Clifford ensemble provides the unitary-design property needed to reconstruct the relevant second- and fourth-order moments from collision statistics of the measured bitstrings.

The half-chain 2-R\'enyi entanglement entropy is defined as $S_2(A,t)\!=\!-\log_2\mathrm{Tr}[\rho_A(t)^2]$, where $\rho_A(t)\!=\!\mathrm{Tr}_{A^c}\rho(t)$ and $A$ is the left half of the chain. We propose to estimate the subsystem purity $P_A(t_k)\!=\!\mathrm{Tr}[\rho_A(t_k)^2]$ from the same bitstrings using the kernel
\begin{align}
K_2(u,v)
&=
(-2)^{-\mathrm{wt}(u\oplus v)} ,
\end{align}
where $\oplus$ denotes bitwise XOR and $\mathrm{wt}(\bullet)$ is the Hamming weight. The estimator takes the U-statistic form
\begin{align}
P_A(t_k)
&=
\frac{1}{N_U}\sum_{m=1}^{N_U}
\frac{1}{N_M(N_M-1)}
\sum_{\alpha\neq\beta}
(-2)^{-\mathrm{wt}\!\Bigl(s^{(m)}_{\alpha,A}\oplus s^{(m)}_{\beta,A}\Bigr)},
\end{align}
where $S_2(A,t_k)\!=\!-\log_2 P_A(t_k)$.
In parallel, we propose to estimate the global purity $P(t_k)=\mathrm{Tr}[\rho(t_k)^2]$ from the same data which is useful both as a diagnostic of decoherence and as an explicit mixedness correction for $M_2$ via
\begin{align}
P(t_k)
&=
\frac{1}{N_U}\sum_{m=1}^{N_U}
\frac{1}{N_M(N_M-1)}
\sum_{\alpha\neq\beta}
(-2)^{-\mathrm{wt}\!\Bigl(s^{(m)}_{\alpha}\oplus s^{(m)}_{\beta}\Bigr)} .
\end{align}

To quantify nonstabilizerness, we propose to use SRE at $k=2$ defined through the Pauli fourth moment. Let $D=2^L$ and let $\mathcal P_L$ denote $L$-qubit Pauli operators up to an overall phase. Define
\begin{align}
W(t_k)
&:=
D^{-2}\sum_{P\in\mathcal P_L}\langle P\rangle_{t_k}^{4},
\quad
\langle P\rangle_{t_k}=\mathrm{Tr}[P\,\rho(t_k)] .
\end{align}
A mixed-state-consistent definition of the $M_2$ is then
\begin{align}
M_2(t_k)
&=
-\log_2\!\Bigl[\frac{D\,W(t_k)}{\mathrm{Tr}(\rho(t_k)^2)}\Bigr] ,
\end{align}
which reduces to $M_2(t_k)\!\approx\! -\log_2[D\,W(t_k)]$ when $\rho(t_k)$ remains close to pure. We propose to estimate $W(t_k)$ using the four-string kernel
\begin{align}
K_4(u_1,u_2,u_3,u_4)
&=
(-2)^{-\mathrm{wt}(u_1\oplus u_2\oplus u_3\oplus u_4)} ,
\end{align}
and the unbiased U-statistic estimator (requiring $N_M\ge 4$)
\begin{align}
&\mathcal{W}(t_k)\nonumber\\
&=
\frac{1}{N_U}\sum_{m=1}^{N_U}
\frac{1}{\binom{N_M}{4}}
\sum_{\alpha<\beta<\gamma<\delta}
(-2)^{-\mathrm{wt}\!\Bigl(
s^{(m)}_{\alpha}\oplus s^{(m)}_{\beta}\oplus s^{(m)}_{\gamma}\oplus s^{(m)}_{\delta}
\Bigr)} .
\end{align}
Combining $\mathcal W(t_k)$ with the experimentally accessible global purity estimate $\mathcal P(t_k)$, we propose the following experimental estimator for the $M_2$:
\begin{align}
\mathcal{M}_2(t_k)
&=
-\log_2\!\Bigl[\frac{D\,\mathcal{W}(t_k)}{P(t_k)}\Bigr] .
\end{align}
In the near-pure regime where $\mathcal P(t_k)\!\simeq\!1$, this expression simplifies to $\mathcal{M}_2(t_k)\!\simeq\!-\log_2[D\,\mathcal W(t_k)]$, while in regimes with appreciable decoherence it provides a direct mixedness correction via the ratio form. Repeating the same protocol for all $t_k$ yields time traces $S_2(A,t)$ and $M_2(t)$ that can be compared on equal footing to elucidate the interplay between entanglement growth and nonstabilizerness generation under transverse-field Ising dynamics with a linear Stark tilt.
We propose to quantify statistical uncertainties by bootstrap resampling over the local Clifford setting index $m$, which separates initial-state sampling fluctuations from finite-shot measurement noise while preserving the estimator structure. The total number of single-shot measurements per time point is $N_{\mathrm{tot}}\!=\!N_U\times N_M$. In resource allocation, increasing $N_U$ primarily suppresses fluctuations due to finite unitary sampling, whereas increasing $N_M$ primarily suppresses within-setting shot noise; we propose to choose $(N_U,N_M)$ by balancing these contributions under an overall measurement-budget constraint, with the additional practical requirement $N_M\!\ge\!4$ for the fourth-order estimator $\mathcal W(t_k)$.

\end{document}